\documentclass[twocolumn,tighten]{aastex63}
\usepackage{amssymb}
\usepackage{amsmath}
\usepackage{array,multirow}
\usepackage{color}
\usepackage{bm}

\newcommand{\gyr}{{\rm{Gyr}}}
\newcommand{\myr}{{\rm{Myr}}}

\newlength{\mysize}

\shorttitle{Cluster Monte Carlo Code}
\shortauthors{}

\begin{document}
\title{Modeling Dense Star Clusters in the Milky Way and Beyond with the Cluster Monte Carlo Code}

\author[0000-0003-4175-8881]{Carl L.~Rodriguez}
\affil{McWilliams Center for Cosmology and Department of Physics, Carnegie Mellon University, Pittsburgh, PA 15213, USA}
\email{carlrodriguez@cmu.edu}

\author[0000-0002-9660-9085]{Newlin C.~Weatherford}
\affil{Department of Physics \& Astronomy, Northwestern University, Evanston, IL 60208, USA}
\affil{Center for Interdisciplinary Exploration \& Research in Astrophysics (CIERA), Northwestern University, Evanston, IL 60208, USA}

\author{Scott C.\ Coughlin}
\affil{Center for Interdisciplinary Exploration \& Research in Astrophysics (CIERA), Northwestern University, Evanston, IL 60208, USA}

\author{Pau Amaro Seoane}
\affil{Universitat Polit{\`e}cnica de Val{\`e}ncia,Camino de Vera, s/n 46022 Valencia. Spain}
\affil{DESY Zeuthen, Platanenallee 6 D-15738 Zeuthen, Germany}
\affil{Institute of Applied Mathematics, Academy of Mathematics and Systems Science, CAS, Beijing 100190, China\\
Kavli Institute for Astronomy and Astrophysics, Beijing 100871, China}
\affil{Zentrum f{\"u}r Astronomie und Astrophysik, TU Berlin, Hardenbergstra{\ss}e 36, 10623 Berlin, Germany}

\author[0000-0001-5228-6598]{Katelyn Breivik}
\affil{Center for Computational Astrophysics, Flatiron Institute, 162 Fifth Ave, New York, NY, 10010, USA}

\author[0000-0002-3680-2684]{Sourav Chatterjee}
\affil{Tata Institute of Fundamental Research, Homi Bhabha Road, Mumbai 400005, India}

\author[0000-0002-7330-027X]{Giacomo Fragione}
\affil{Department of Physics \& Astronomy, Northwestern University, Evanston, IL 60208, USA}
\affil{Center for Interdisciplinary Exploration \& Research in Astrophysics (CIERA), Northwestern University, Evanston, IL 60208, USA}

\author[0000-0003-4412-2176]{Fulya K{\i}ro\u{g}lu}
\affil{Department of Physics \& Astronomy, Northwestern University, Evanston, IL 60208, USA}
\affil{Center for Interdisciplinary Exploration \& Research in Astrophysics (CIERA), Northwestern University, Evanston, IL 60208, USA}

\author[0000-0002-4086-3180]{Kyle Kremer}
\affil{TAPIR, California Institute of Technology, Pasadena, CA 91125, USA}
\affil{The Observatories of the Carnegie Institution for Science, Pasadena, CA 91101, USA}

\author[0000-0002-1884-3992]{Nicholas Z.~Rui}
\affil{TAPIR, California Institute of Technology, Pasadena, CA 91125, USA}


\author[0000-0001-9582-881X]{Claire S.~Ye}
\affil{Department of Physics \& Astronomy, Northwestern University, Evanston, IL 60208, USA}
\affil{Center for Interdisciplinary Exploration \& Research in Astrophysics (CIERA), Northwestern University, Evanston, IL 60208, USA}

\author[0000-0002-0147-0835]{Michael Zevin}
\affiliation{Kavli Institute for Cosmological Physics, The University of Chicago, 5640 South Ellis Avenue, Chicago, IL 60637, USA}
\affiliation{Enrico Fermi Institute, The University of Chicago, 933 East 56th Street, Chicago, IL 60637, USA}

\author[0000-0002-7132-418X]{Frederic A.~Rasio}
\affil{Department of Physics \& Astronomy, Northwestern University, Evanston, IL 60208, USA}
\affil{Center for Interdisciplinary Exploration \& Research in Astrophysics (CIERA), Northwestern University, Evanston, IL 60208, USA}

\begin{abstract}
We describe the public release of the Cluster Monte Carlo Code (\texttt{CMC}) a parallel, star-by-star $N$-body code for modeling dense star clusters.  \texttt{CMC} treats collisional stellar dynamics using H\'enon's method, where the cumulative effect of many two-body encounters is statistically reproduced as a single effective encounter between nearest-neighbor particles on a relaxation timescale.  The star-by-star approach allows for the inclusion of additional physics, including strong gravitational three- and four-body encounters, two-body tidal and gravitational-wave captures, mass loss in arbitrary galactic tidal fields, and stellar evolution for both single and binary stars.  The public release of \texttt{CMC} is pinned directly to the \texttt{COSMIC} population synthesis code, allowing dynamical star cluster simulations and population synthesis studies to be performed using identical assumptions about the stellar physics and initial conditions.  As a demonstration, we present two examples of star cluster modeling: first, we perform the largest ($N=10^8$) star-by-star $N$-body simulation of a Plummer sphere evolving to core collapse, reproducing the expected self-similar density profile over more than 15 orders of magnitude; second, we generate realistic models for typical globular clusters, and we show that their dynamical evolution can produce significant numbers of black hole mergers with masses greater than those produced from isolated binary evolution (such as GW190521, a recently reported merger with component masses in the pulsational pair-instability mass gap).  
\end{abstract}


\section{Introduction} \label{S:intro}

The modeling of dense star clusters (DSCs), such as globular clusters (GCs), super star clusters (SSCs), or the nuclear star clusters (NSCs) in the centers of many galaxies, remains one of the most challenging problems in computational astrophysics.  At first glance, this fact is somewhat surprising: these clusters typically start with $10^5-10^8$ individual particles, well within the range of many collisionless gravitational particle solvers.  However, it is the \emph{collisional} nature of these star clusters that makes their dynamics so difficult to resolve: unlike the dynamics of galaxies or particles in a cosmological volume, the long-term evolution of DSCs is driven by both the orbits of individual stars in the cluster potential and the diffusion of the stars through phase space via two-body encounters \cite[e.g.,][]{Spitzer1987,Heggie2003,Binney2008}.  

Both historically and presently, the first approach employed to study collisional stellar dynamics has often been a direct summation approach.  In many ways the most straightforward conceptually (though exceedingly complex in implementation), a direct summation code calculates the gravitational force of every particle on every other particle before summing the individual forces to arrive at the instantaneous acceleration for every particle.  The system is then advanced using standard numerical techniques for solving ordinary differential equations (typically a fourth-order Hermite integrator).  These techniques have a long and storied history going back more than 60 years to the first numerical work by \cite{VonHoerner1960}.  The modern generation of codes, such as \texttt{HiGPUs} \citep{Capuzzo-Dolcetta2013a}, \texttt{PhiGRAPE} \citep{2008MNRAS.389....2H}, \texttt{ph4} \citep{McMillan2012}, \texttt{frost} \citep{2021MNRAS.502.5546R}, and the well-known \texttt{NBODY} series \citep{Aarseth2003,Aarseth2012} and its derivatives \citep[\texttt{NBODY6++GPU},][]{Wang2015} contain state-of-the-art algorithmic and hardware optimizations that enable the modeling of clusters with $N\sim10^6$ particles.  But even with dozens of parallel GPUs, these models require months or even years of wall-clock time to integrate a cluster for $\sim10\rm{Gyr}$, \cite[e.g.,][]{Heggie2014,Wang2016,2021MNRAS.502.5546R}, and are limited to systems with low binary fractions, large initial radii, and relatively long dynamical times.  These constraints preclude any reasonable exploration of the parameter space of massive star clusters, especially those with the compact initial radii needed to produce core-collapsed GCs \cite[e.g.][]{Kremer2019a}.   

Direct summation $N$-body methods represent the ``gold standard'' of collisional stellar dynamics because of their unparalleled accuracy.   But just as the advent of fiat currency proved faster, cheaper, and more robust than the gold standard, faster approaches to collisional stellar dynamics offer many avenues of scientific study.  The orbit-averaged Monte Carlo method, originally developed by \cite{Henon1971a,Henon1971b}, is one such approach.  Instead of integrating the orbits of each star directly, the Monte Carlo method leverages a statistical treatment of stellar dynamics, where the cumulative effect of many distant two-body encounters is modeled as a single effective scattering between neighboring particles.  The deflection angle of this effective encounter is chosen to reproduce the mean of the averaged change in the velocity per unit time, $\left<(\Delta v)\right>^2$, experienced by a particle traveling through a field of stars with known density \cite[e.g.,][Chapter 2]{Spitzer1987}.  In doing so, H\'enon's method naturally resolves the two-body relaxation that drives the long-term evolution of collisional star systems, and has been shown to reproduce the pre- and post-core collapse evolution of DSCs for nearly 50 years \citep{Aarseth1974,Joshi2000,2013MNRAS.431.2184G,Rodriguez2016b, Kremer2020}.  Furthermore, by modeling collisional dynamics as nearest-neighbor encounters between radially-sorted particles, the Monte Carlo method can also be expanded to include close encounters between stars and binaries, a critical component to understanding the post-collapse evolution of clusters and their production of stellar exotica such as X-ray binaries \citep{Clark1975,Pooley2003,Kremer2018a}, millisecond pulsars \citep{Rappaport1989, Ye2019}, cataclysmic variables \citep{1995ApJ...455L..47G,Kremer2021}, and binary black hole (BBH) mergers \citep{PortegiesZwart2000,Rodriguez2015a}. 

{Following updates to the method by \cite{Stodoikiewicz1982} to ensure the stability of long-term integrations, the majority of work on the H\'enon $N$-body method has been led by two groups: the Cluster Monte Carlo (\texttt{CMC}) group \citep{Joshi2000} and the Monte Carlo Cluster Simulator (\texttt{MOCCA}) group \citep{Giersz1998}.  A third code was also developed around the same time by \cite{Freitag2001} which introduced new ways of treating stellar collisions and central-massive BHs with H\'enon's method.  While that code is no longer actively maintained, many of its techniques have since been incorporated into both \texttt{CMC} and \texttt{MOCCA}.  Both codes have seen significant enhancements and improvements over the last two decades, with the most recent iterations of the codes being described in \cite{Pattabiraman2013} and \cite{Giersz2013}.  Although there are key differences in the choices of the dynamical timestep and other aspects of stellar evolution and collisions, \texttt{CMC} and \texttt{MOCCA} contain much of the same physics, and are currently the only codes capable of modeling realistic populations of DSCs with more than $10^6$ stars and binaries over many Gyr.}

In this paper, we describe the first public release of the Cluster Monte Carlo code, \texttt{CMC}, an $N$-body approach to modeling collisional stellar dynamics with H\'enon's method.  Written in \texttt{C} and designed for distributed-memory architectures, \texttt{CMC} has been continuously developed since \cite{Joshi2000}, and contains all the necessary physics for the evolution of massive, spherical DSCs.  Key physical processes, such as two- and three-body binary formation, strong encounters between stars and binaries \citep[performed with the small-$N$ direct integrator \texttt{fewbody},][]{Fregeau2003,Fregeau2007}, galactic tidal fields \citep{Joshi2001a,RodriguezInPrep}, physical collisions \citep{Fregeau2007} and post Newtonian dynamics \citep{Rodriguez2018} have been incorporated and are available in the v1.0 release of \texttt{CMC} \citep{codepaper2021}. 

The public release of \texttt{CMC} also contains detailed treatments for single and binary stellar evolution using the \texttt{COSMIC} population synthesis code \citep{Breivik2020}.  Both \texttt{CMC} and \texttt{COSMIC} are based on the Binary Stellar Evolution (\texttt{BSE}) package of \cite{Hurley2000,Hurley2002}, and have been significantly enhanced with new prescriptions for the evolution of massive stars, binary mass transfer, common-envelope efficiency, compact object formation, and much more.  The latest version of \texttt{COSMIC}, v3.4, is now pinned as a Git submodule to \texttt{CMC}, ensuring that any changes to the population synthesis code are also available in the dynamics code.  This allows for direct, ``apples-to-apples'' comparisons between isolated binary evolution and dynamical formation for many astrophysical sources, such as stellar binaries and high-energy transients.  In Section \ref{S:code}, we review the H\'enon method, its implementation in \texttt{CMC} and the additional dynamical and stellar physics included in the code, as well as the assumptions and limitations of the Monte Carlo method.  In Section \ref{S:release}, we describe features specific to the public release of \texttt{CMC}, and the integration of the \texttt{COSMIC} population synthesis package.  Finally in Section \ref{S:examples}, we show two examples of cluster evolution: a Plummer sphere of point-mass particles evolved to core collapse, and several realistic cluster models evolved for 12 Gyr.




\section{Dynamics in \texttt{CMC}} \label{S:code}
Here we review the theoretical basis of the original \cite{Henon1971a,Henon1971b} scheme as it is implemented in \texttt{CMC}.  In a collisional stellar system, particles undergo repeated two-body gravitational encounters with other particles.  These ``collisions'' allow energy and angular momentum to be exchanged between particles and subsequently diffused throughout the cluster, just as physical collisions diffuse energy throughout an ideal gas (though, unlike their microscopic counterparts, a star cluster can never reach thermodynamic equilibrium).  In clusters with relatively few particles ($N \lesssim 10^3$), this process is dominated by a handful of close encounters with small impact parameters, producing large deflection angles with every encounter.  With larger $N$, however, the cumulative effect of many distant encounters (producing many small-angle deflections in the particle's trajectory) begin to dominate the cluster evolution \cite[e.g.,][]{Spitzer1987}.  In the large-$N$ regime, the dynamical behavior of the particles becomes largely predictable using the techniques of statistical mechanics, forming the basis for the H\'enon method.  

The dividing line between these regimes can be understood as a competition between two timescales: the \emph{dynamical} time (the time a particle takes to orbit the cluster at the half-mass radius), and the \emph{relaxation} timescale, or the time required for these distant encounters to change the velocity of the particle by order of itself.  The scaling between these two timescales depends on the size of the cluster \cite[e.g.,][]{Binney2008}:

\begin{equation}
T_{\rm relax} \approx \frac{0.1 N}{\log N} T_{\rm dyn}~.
\label{eqn:approxtrel}
\end{equation}

\noindent When $T_{\rm relax} \gg T_{\rm dyn}$, the cluster is in the weak-encounter regime, and can be thought of as a collection of stars on semi-independent orbits that slowly diffuse through phase space on a relaxation timescale.  

By itself, two-body relaxation would drive star clusters towards core collapse, where the inner regions of the cluster contract as energy is transported by two-body relaxation from the core towards the outer halo of the cluster.  This, in turn, causes an expansion of the outer regions of the halo, and creates a dynamical ``temperature'' gradient between the hotter core and the cooler outer regions.  This gradient accelerates the transport of heat from core to halo, driving the central regions of the cluster inwards and towards the well-known gravothermal catastrophe \cite[e.g.,][]{1985IAUS..113..525A,1968MNRAS.138..495L,Heggie2003}.  But while these distant two-body encounters drive the cluster to collapse, it is the close encounters that largely reverse this process.  As the core contracts, the increasing density of objects facilitates the production and dynamical hardening of binaries, creating an effective power source that halts the collapse and determines the long-term evolutionary fate of the cluster.  These close encounters are also largely responsible for the large number of stellar exotica found in GCs and other DSCs.

In \texttt{CMC}, the dynamics of inter-particle interactions are separated into these two natural categories: (a) the average two-body relaxation from many weak encounters using H\'enon's method and (b) strong encounters, such as binary encounters and physical collisions.  We now describe both of these.  
\
\

\subsection{Two-Body Relaxation} \label{S:relaxation}

  \begin{figure*}
\centering
\includegraphics[]{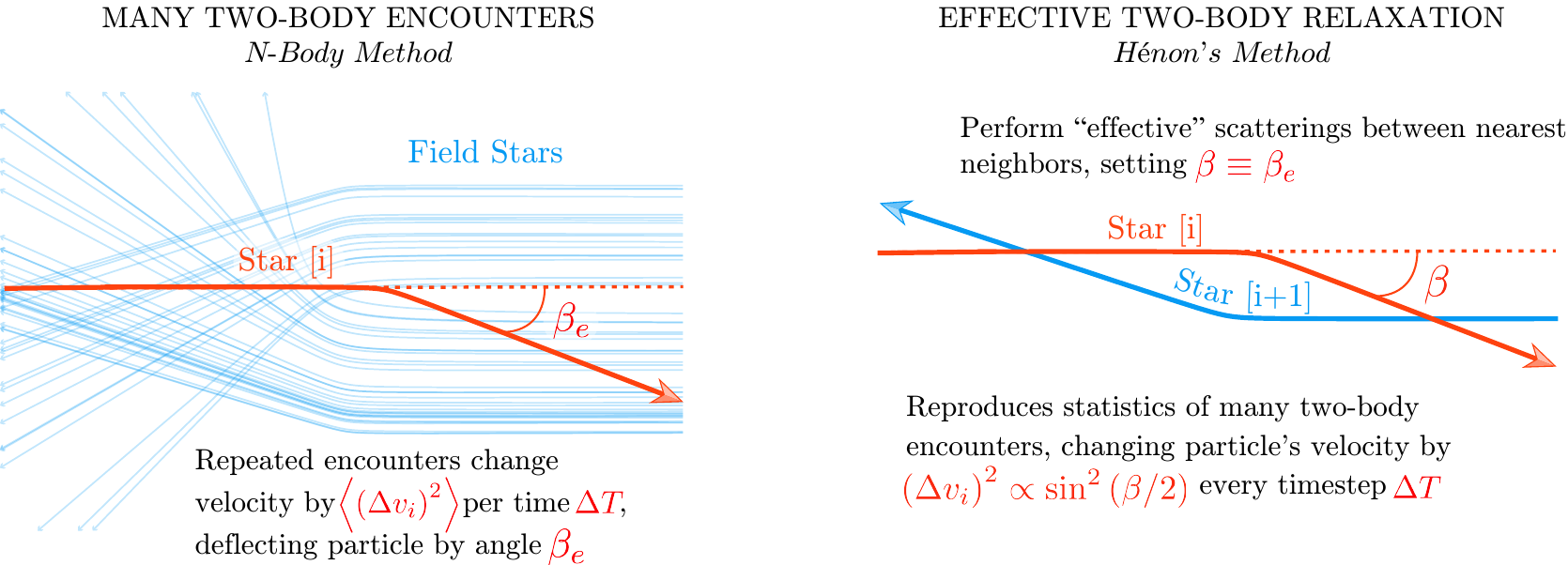}
\caption{Schematic diagram of the H\'enon method.  The cumulative velocity change due to many two-body encounters in a realistic cluster can be calculated given the average local mass, velocity, and density of stars with equation \eqref{Eq:relax4}.   In H\'enon's method, this is reduced to a single encounter between nearest neighbor particles, where the deflection angle is chosen to reproduce the statistical predictions using  equation \eqref{Eq:relax6}.}
\label{fig:scattering}
\end{figure*}

In the H\'enon method, the cluster is assumed to be spherically symmetric and in dynamical (virial) equilibrium.  The full 6-D positions and velocities are reduced to 3-D: the radial position $r$ of each particle and its radial and tangential velocities, $v_r$ and $v_t$.  Of course, these position and velocities change on a dynamical timescale as the particle orbits through the cluster potential.  What the Monte Carlo method tracks over time is the energy ($E$) and angular momentum ($J$) of each particle's orbit, quantities that are conserved on dynamical timescales.  Combined with the cluster potential, $\Phi(r)$, these quantities fully define an orbit for every particle in the cluster.  New $r$, $v_r$, and $v_t$ are randomly sampled from these orbits at every Monte Carlo timestep.

While $E$ and $J$ do not change on dynamical timescales, the cumulative effect of many weak encounters drives their evolution on a relaxation timescale.  During a Monte Carlo timestep, we model these weak encounters as perturbations $\Delta E$ and $\Delta J$ to the particle's energy and angular momentum. These perturbations are computed as a single \textit{effective} encounter between each particle and its nearest neighbor in radius. In the center-of-mass frame of the two-body encounter, the \emph{magnitude} of the particles' velocities is unchanged, but the velocity vectors are deflected by some angle $\beta$. In this frame, the energy of the two particles is conserved; however, in the \emph{cluster} frame, such encounters exchange energy and angular momentum. To correctly capture two-body relaxation throughout the cluster, each of these effective encounters should give the correct value for the mean change in kinetic energy at each particle's position during the timestep $\Delta T$. We satisfy this constraint by computing an effective deflection angle $\beta_e$ for each particle that results in the correct mean $\Delta E$ and $\Delta J$.  See Figure~\ref{fig:scattering}.

\subsubsection{Calculating the effective scattering angle}
\label{sec:beta}

We now proceed to derive the effective scattering angle, $\beta_e$, following the logic of \citet{Stodoikiewicz1982} and \citet{Joshi2000}.  We begin with an array of particles sorted by increasing radius from the cluster center such that $r_i < r_{i+1}$.  Fundamentally we are interested in the average change in energy and angular momentum, $\Delta E$ and $\Delta J$, particle $i$ experiences during some time interval $\Delta T$.  We begin by considering the change in velocity squared, $(\Delta v_i)^2$, for a single two-body interaction between the star with index $i$ and a field star with mass $m_f$.  This expression can be written in terms of $\beta$, the angle of deflection of particle $i$, or in terms of $b$, the impact parameter of the encounter at infinity, as \cite[e.g.,][Eq.~3.54]{Binney2008}
\begin{align} 
(\Delta v_i)^2 & = \frac{4\,m_{f}^2}{(m_i+m_{f})^2} w^2\sin^2 \left(\frac{\beta}{2}\right) \label{Eq:relax2}\\
 & = \frac{4\,G^2\,m_{f}^2}{w^2\,b_{90}^2} \frac{1}{1+(b/b_{90})^2}, \label{Eq:relax25}
\end{align}
where $w = |\mathbf{v}_i - \mathbf{v}_{f}|$ is the relative speed of the particles at infinity and $b_{90} \equiv G(m_i+m_{f})/w^2$ is the impact parameter that would deflect the particles by $\beta = 90^{\circ}$. 

It is straightforward to extend equation \eqref{Eq:relax25} to describe the effect of many encounters.  The change in velocity over time due to many weak encounters can be calculated by considering the field particles encountered by particle $i$ as it travels through the cluster. The number of particles encountered at an impact parameter $b$ during an interval $\Delta T$ is simply the product of the local number density of particles, $n$, and the volume swept out at that impact parameter.   For an infinitesimal annulus with inner radius $b$ and outer radius $b+db$, this product is $N_{\rm enc}=2\pi b n w \Delta T db$, where the length of the cylinder is the relative velocity of the particles, $w$, times $\Delta T$. To compute the contribution of this result over all impact parameters, we multiply equation \eqref{Eq:relax25} by $N_{\rm enc}$ and integrate over $b$ from zero to some upper limit $b_{\rm{max}}$:

\begin{align}
\left<(\Delta v_i)^2\right> & = \frac{8\pi G^2 n \Delta T\,m_{f}^2}{w b_{90}^2} \int_0^{b_{\rm max}}\frac{b\cdot db}{1+(b/b_{90})^2} \label{eq:int} \\
& = 4 \pi G^2 n \Delta T\,m_{f}^2 w^{-1} \ln \left[1+(b_{\rm max}/b_{90})^2\right] \label{Eq:relax3}\\
&\simeq 8 \pi G^2 n \Delta T\,m_{f}^2 w^{-1} \ln \Lambda~.  \label{Eq:relax35}
\end{align}

The value of $b_{\rm max}$ to use in equation \eqref{eq:int} is a well-studied complication to the theory: on the one hand, in a system with uniform density, equal octaves of $b$ contribute equally to $\left<(\Delta v_i)^2\right>$, meaning for sufficiently large systems, $b_{\rm max} \gg b_{90}$; on the other hand, star clusters are finite systems whose densities eventually fall to zero at some finite radii.  In going from equation \eqref{Eq:relax3} to equation \eqref{Eq:relax35}, we have introduced the well-known Coulomb logarithm, defined as

\begin{equation}
\Lambda \equiv \frac{b_{\rm{max}}}{b_{90}} = \frac{w^2 b_{\rm max}}{G (m_i+m_f)}~,
\end{equation}

\noindent and used our assumption of the weak encounter regime to set $\ln\left[1+(b_{\rm max}/b_{90})^2\right] \simeq 2\ln\left[b_{\rm max}/b_{90}\right]$.  $\Lambda$ can be approximately computed at the half-mass radius of the cluster by equating $b_{\rm max} = r_h$ and replacing $w^2$ with the mean-squared velocity from the virial theorem, $\left<v^2\right> \simeq 0.45 G m_f /r_h$ \cite[][p.~361]{Binney2008}.  For a cluster with $N$ stars of mass $m_f$, this yields

\begin{equation}
\label{eqn:loglambda}
\ln \Lambda = \ln\left[\gamma N\right]~,
\end{equation}

\noindent with $\gamma \simeq 0.2$.  However, comparisons to direct $N$-body simulations have suggested a $\gamma$ of 0.11 \cite[for equal-mass clusters, e.g.,][]{1994MNRAS.268..257G}, 0.02 \cite[for clusters with realistic mass functions,][]{1996MNRAS.279.1037G} or even 0.01 \cite[the current default in \texttt{CMC},][]{Freitag2006b,Rodriguez2015c}, to be more realistic.  

Equations (\ref{eq:int}-\ref{Eq:relax35}) were computed assuming that every field star has the same mass and relative velocity, $m_f$ and $w$, with respect to our test star, $m_i$.  In realistic clusters, however, nearby particles will have a range of masses and velocities, determined by the distribution function, $F(\mathbf{r},\mathbf{v},m)$.  Therefore, to accurately reproduce the cumulative shift $\left<(\Delta v_i)^2\right>$ from equation \eqref{Eq:relax35}, we must replace the $m_{f}^2 w^{-1}$ factor with

\begin{equation} \label{Eq:relax4}
\left<(\Delta v_i)^2\right> = 8 \pi G^2 n \Delta T \ln\Lambda \left< m_{f}^2 w^{-1} \right>_{F},
\end{equation}

\noindent where we use $\left< ... \right>_{F}$ to indicate an average over the local phase-space distribution function

\begin{equation}
\label{argh2}
\left< m_{f}^2 w^{-1} \right>_{F} \equiv \int F_i F_f m_f^2 w^{-1} d^3\mathbf{v}_i d^3\mathbf{v}_f dm_i dm_f
\end{equation}

\noindent with $F_i = F(\mathbf{r}_i,\mathbf{v}_i,m_i)$ and $F_f = F(\mathbf{r}_f,\mathbf{v}_f,m_f)$. 

We now reach the crux of the Monte Carlo method: to get the angle of deflection $\beta$ in terms of these quantities -- and the correct mean value of $m_i(\Delta v_i)^2$ -- we equate the right-hand side of equation \eqref{Eq:relax4} to the right-hand side of equation \eqref{Eq:relax2}:
\begin{align} 
\left<\frac{4\,m_{f}^2\,w^2}{(m_i+m_{f})^2}\right>_F &\sin^2 \left(\frac{\beta_e}{2}\right) \label{Eq:relax5}\\ &=8 \pi G^2 n \Delta T \ln \Lambda \left< m_{f}^2 w^{-1}\right>_F,\nonumber
\end{align}
where we have replaced a single deflection angle, $\beta$, with an effective deflection angle, $\beta_e$, and replaced the mass and velocity terms with their averaged equivalents.  We now solve for $\beta_e$, which involves averaging the mass and velocity quantities over the relevant distribution functions.  However, as \cite{Henon1971a} pointed out, the most straightforward way to sample the distribution function near star $m_i$ is to pick the mass and velocity of its nearest neighbor, $m_{i+1}$.  While this is not the same as a true average, the nearest neighbor represents a fair draw from the mass and velocity distributions, and would reproduce the true distribution function after a sufficient number of timesteps.  Dispensing with the distribution-function averages and substituting $m_f\rightarrow m_{i+1}$, equation \eqref{Eq:relax5} becomes
\begin{equation} \label{Eq:relax6}
\sin^2 \biggr( \frac{\beta_e}{2}\biggr) = \frac{2\pi G^2 (m_i+m_{i+ 1})^2}{w^3} n \ln \Lambda \Delta T.
\end{equation}

\noindent The only quantity that must be explicitly averaged is the local volumetric density of particles, $n$.  In \texttt{CMC}, this is done with a moving average over 41 stars, from $i-20$ to $i+20$ for each star $i$.

For computational convenience, equation \eqref{Eq:relax6} is more easily expressed in terms of ``relaxation timescale'' between particles $i$ and $i+1$:

\begin{equation} \label{Eq:relax8}
\sin \biggr(\frac{\beta_e}{2}\biggr) = \frac{\pi}{4} \sqrt{\frac{\Delta T}{T_{{\rm rel},i}}}~,
\end{equation}

\noindent where

\begin{equation}  \label{Eq:relax7}
T_{{\rm rel},i} \equiv \frac{\pi}{32} \frac{w^3}{G^2 n (m_i + m_{i+1})^2 \ln\Lambda}
\end{equation}

\noindent is the typical time required for a particle to be deflected by $90^{\circ} $ \citep{Freitag2001}.  Note that, in practice, we clamp the maximum $\beta_e$ between any two particles at $90^{\circ}$.

\subsubsection{Calculating new $E$ and $J$}
\label{sec:coord}

With $\beta_e$ computed for each pair of particles, we now convert these pairwise deflection angles into $\Delta E$ and $\Delta J$ for each orbit.  We follow H\'enon's original approach, placing our two-body encounter into 3-D space.  We denote the phase space coordinates of the two interacting particles by $(r_i,v_{r,i},v_{t,i})$ and $(r_{i+1},v_{r,i+1},v_{t,i+1})$, with masses $m_i$ and $m_{i+1}$, respectively. We choose our reference frame  such that the $z$-axis is parallel to $\bm{r}_i$ and the $(x,z)$-plane contains $\bm{v}_i$. In Cartesian coordinates, the two particle velocities are then
\begin{align} 
\bm{v}_i&=(v_{t,i},0,v_{r,i}),\nonumber\\
\bm{v}_{i+1}&=(v_{t,i+1}\cos\phi,v_{t,i+1}\sin\phi,v_{r,i+1}), \label{Eq:orbits1}
\end{align}
where $\phi$ is a uniform random variate in the range $[0,2\pi]$, since spherical symmetry guarantees isotropic transverse velocities. The relative velocity is then:

\begin{align} \label{Eq:orbits2}
\bm{w}\equiv&\bm{v}_{i+1}-\bm{v}_i\nonumber\\
=&(v_{t,i+1}\cos\phi-v_{i,t},~ v_{t,i+1}\sin\phi,~v_{r,i+1}-v_{r,i}).
\end{align}

We now define vectors $\bm{w_1}$ and $\bm{w_2}$ with magnitude equal to $|\bm{w}|$, such that $\bm{w_1} \times \bm{w_2} = \bm{w}$. In this right-handed coordinate system, we have
\begin{align} \label{Eq:orbits3}
\bm{w_1}\equiv&\Bigr(\frac{-w_y w}{w_p},\frac{w_x w}{w_p},0\Bigr),\nonumber\\
\bm{w_2}\equiv&\Bigr(\frac{-w_x w_z}{w_p}, \frac{-w_y w_z}{w_p}, w_p\Bigr),
\end{align}
where $w_p=\sqrt{w_x^2+w_y^2}$. Since the distribution of field stars is assumed to be spherically symmetric, we randomly select the angle $\psi\in[0,2\pi]$ between the plane of relative motion, defined by ($\bm{r}_{i+1}-\bm{r}_i,\bm{v}_{i+1}-\bm{v}_i)$, and the plane containing $\bm{w}$ and $\bm{w_1}$. The relative velocity after the encounter, $\bm{w}^{\rm new}$, is then
\begin{equation} \label{Eq:orbits4}
\bm{w}^{\rm new} = \bm{w}\cos\beta_e + \bm{w_1}\sin\beta_e\cos\psi + \bm{w_2}\sin\beta_e\sin\psi,
\end{equation}
where $\beta_e$ is the effective scattering angle of Section \ref{S:relaxation}. The post-interaction particle velocities in the cluster frame, $\bm{v}^{\rm new}_i$ and $\bm{v}^{\rm new}_{i+1}$, become:

\begin{align} 
\bm{v}^{\rm new}_{i} &= \bm{v}_i-\left(\frac{m_{i+1}}{m_i+m_{i+1}}\right)(\bm{w}^{\rm new}-\bm{w}),\nonumber\\ 
\bm{v}^{\rm new}_{i+1} &= \bm{v}_{i+1}+\left(\frac{m_{i}}{m_i+m_{i+1}}\right)(\bm{w}^{\rm new}-\bm{w}).
\end{align}

Since we set the $z$-axis parallel to $\bm{r}^i$, the new radial and transverse velocities for the first particle are $v_{r,i}^{\rm new}=v^{\rm new}_{z,i}$ and $v^{\rm new}_{t,i}=\sqrt{(v^{\rm new}_{x,i})^2+(v^{\rm new}_{y,i})^2}$. The new specific orbital energy and specific angular momentum are then $E_i=\Phi(r_i)+\frac{1}{2}\left(\left(v^{\rm new}_{r,i}\right)^2+\left(v^{\rm new}_{t,i}\right)^2\right)$ and $J_i=r_i v^{\rm new}_{t,i}$, respectively, and similarly for the $i+1$ particle.


\subsection{Strong Interactions} \label{S:stronginteractions}

Layered onto the basic two-body relaxation calculation is the possibility for strong interactions to occur between neighboring particles at each timestep. Strong interactions can be separated into two groups: single--single interactions (which include physical collisions, tidal disruptions, tidal captures, and gravitational-wave captures of pairs of BHs) and binary interactions involving one or two binaries (integrated directly using \texttt{fewbody}). Strong interactions are sampled following the method of \citet{Freitag2002}---see also \citet{Fregeau2007}---where we evaluate the quantity

\begin{equation}
    \label{Eq:strong_enc}
    P_{\rm{strong}} = n w \Sigma_{\rm{strong}} \Delta T
\end{equation}
between each pair of neighboring stars and binaries. Here, $P_{\rm{strong}}$ is the probability for a strong interaction with cross-section $\Sigma_{\rm{strong}}$ to occur, $n$ is the local number density of stars, $w$ is the relative velocity of the pair of objects at infinity (computed in the same coordinate system as Section \ref{sec:coord}), and $\Delta T$ is the timestep.  Accounting for gravitational focusing, $\Sigma_{\rm{strong}}$ can be written as

\begin{equation}
    \label{Eq:sigma_strong}
    \Sigma_{\rm{strong}} = \pi r_p^2 \left(1 + \frac{2GM}{r_p w^2} \right) ~,
\end{equation}
where $M$ is the total mass of the pair of interacting objects and $r_p$ is the maximum distance of pericenter along a hyperbolic for each strong encounter to occur.  At each timestep, a random number ${X}$ is drawn from a uniform distribution between $0$ and $1$.  If ${X}<P_{\rm{strong}}$, a strong interaction occurs; otherwise, the particles undergo standard two-body relaxation.

The value of $r_p$ is determined by the type of interaction, as described in following subsections.  We note that both the strong encounters described here and the binary formation considered in Section \ref{sec:binaries} are actually performed \emph{before} two-body relaxation is considered in \texttt{CMC}.  Any particles that participate in strong encounters do not participate in two-body relaxation during the same timestep.  
{This is because, on a relaxation timescale, the change in a particle's orbit due to a strong encounter occurs instantaneously, and is often significantly greater than the gradual changes due to two-body encounters.  These large changes in velocity render the calculation in Section \ref{sec:beta} invalid, since the relative velocity between field and test stars is no longer constant over a given time interval.  However, because our timestep is chosen as the minimum of the relaxation and strong encounter timescales (the latter of which is typically much smaller than the former), the deflection by two-body relaxation during a timestep where a strong encounter occurs is typically minimal.  See Section \ref{S:timestep}.}

\subsubsection{Single--Single Collisions}
\label{sec:sscoll}

For collisions between single stars, \texttt{CMC} operates under the sticky-sphere approximation, where any two particles that touch radii are assumed to have collided.  The cross-section for a pair of single stars with masses $m_i$ and $m_{i+1}$ and radii $R_i$ and $R_{i+1}$ can be expressed as

\begin{equation}
    \Sigma_{\rm{coll}} = \pi (R_i+R_{i+1})^2 \left(1 + \frac{2GM}{(R_i+R_{i+1}) w^2} \right)~,
\end{equation}
where $M=m_i+m_{i+1}$.

For main-sequence and/or giant stars, we adopt the classic sticky-sphere approximation, where zero mass is lost during the collision.  The mass of the collision product, $m_{\rm new}$, is simply computed as $m_i+m_{i+1}$, with the new evolutionary phase of the star determined self-consistently using the stellar evolution prescriptions in \texttt{COSMIC}. In GCs where the relative velocity of a pair of stars is typically much less than the escape velocity at the surface of the star, this approximation is reasonable \citep[e.g.,][]{Lombardi2002}.  For compact objects, however, we employ more conservative assumptions for the mass retained by merger products. For collisions of white dwarfs with other white dwarfs, main sequence stars, or giants, we follow the prescriptions described in \citet[][Section 2.7]{Hurley2002}.  As described in \citet{Kremer2019b}, when a main-sequence star or giant collides with a BH/NS, we assume the accretion feedback supplies sufficient energy to unbind the stellar material from the system completely.  In this case, negligible mass is accreted and the mass of the BH/NS is unchanged after the collision.  {We note that this is an extremely conservative assumption, particularly if the star is significantly more massive than the BH or NS.  However, in practice such collisions are rare in most typical GC models; see Section \ref{subsec:bbh}.}

\subsubsection{Binary Interactions}
\label{binint}

The maximum pericenter distance at which a dynamical binary interaction may occur is given by $r_p=\mathcal{X}_{\rm{bin}}a$ (for three-body encounters between a star and a binary with semi-major axis $a$) or $r_p=\mathcal{X}_{\rm{bin}}(a_1+a_2)$ (for four-body encounters between two binaries with semi-major axes $a_1$ and $a_2$). Here, $\mathcal{X}_{\rm{bin}}$ is a free parameter chosen so that all binary encounters of interest (i.e., those encounters which are energy-generating) are captured. An arbitrarily large value of $\mathcal{X}_{\rm{bin}}$ would capture all binary encounters (the majority of which will be weak flybys); however, this would come at increased computational cost, as the global timestep would shrink significantly to resolve these encounters (see Section \ref{S:timestep}). As described in \citet{Fregeau2007}, we adopt a fiducial value of $\mathcal{X}_{\rm{bin}}=2$, a compromise between capturing all possible encounters and computational cost.  {We note that this choice limits the number of weak flyby encounters that can occur in the cluster.  While these encounters do not typically change the energy of the binaries (and are therefore irrelevant for the long-term evolution of the cluster), they can slowly alter the binary angular momentum, driving the system to higher eccentricites that will eventually result in mass transfer and even mergers \citep[e.g.,][]{1995ApJ...445L.133R,2019MNRAS.487.5630H}.  Efforts to model this in \texttt{CMC} are currently underway.}

{For binary-single encounters, we assume that the test star is a binary, and calculate the probability for it to encounter a single object following equation \eqref{Eq:strong_enc}.  For binary-binary encounters, we should formally replace the $n$ in equation \eqref{Eq:strong_enc} with $n_{\rm binary}$.  However, the probability for binary-binary encounters to occur is only considered whenever the field star is also a binary, and the probability of \emph{that} being true for any given particle is proportional to the density of binaries over the density of all objects, or $n_{\rm binary} / n$.  By allowing binary-binary encounters to occur only for neighboring binaries, the probability from equation \eqref{Eq:strong_enc} is implicitly multiplied by $n_{\rm binary} / n$ in true Monte Carlo fashion.}

Binary encounters are integrated directly using the small-$N$ scattering code, \texttt{fewbody}.  These encounters depend on the relative velocity, $w$, of the pair of interacting objects (singles or binaries) and the impact parameter at infinity, $b$, which is sampled uniformly in annulus from a disk with maximum radius set by equation \eqref{Eq:sigma_strong}. All interactions are integrated until they reach either an unambiguous endstate or a maximum integration time (either $10^6$ dynamical times of the encounter or 1 minute of CPU wall-clock time).  Encounters that do not resolve into steady systems within these limits are discarded and treated as a standard two-body relaxation.  See \cite{Fregeau2007} for details. 

The version of \texttt{fewbody} in the public release of \texttt{CMC} also contains post Newtonian (pN) corrections to the equations of motion, including the 2.5pN term responsible for GW emission during close encounters \citep{Antognini2014,Amaro-Seoane2016,Rodriguez2018,Rodriguez2018c}.  By default, the 2.5pN term is included in the equations of motion when at least two BHs are present in an encounter.  Note that, to accommodate the additional integration time required to resolve the inspiral and merger of BHs, the wall-clock time limit for these encounters is increased to 10 minutes when pN terms are included.   {We do not include the 1 and 2pN terms in the integration by default, since they make energy conservation within the encounters impossible to track with the standard implementation of \texttt{fewbody} \cite[see][Figure 2, top panel]{Rodriguez2018c}, and have been shown to have no impact on the distribution of merging BBHs eccentricities \citep{2019ApJ...871...91Z}. }

One possible outcome of a binary--binary encounter is a stable hierarchical triple \citep[e.g.,][]{Mikkola1983,Rasio1995,Fragione2020,MartinezFragione2020}. As described in \citet{Fregeau2007}, if such a triple forms within a \texttt{CMC} simulation, we break the triple by unbinding the single star and inner binary, distributing its energy to nearby particles in the cluster.  Additionally, because these \texttt{fewbody} encounters are performed in isolation (that is, without tidal forces from nearby particles), they can form pathologically-wide binaries with separations larger than the typical inter-particle distance.  We destroy any such binaries (defined as having semi-major axes greater than 10\% of the average local inter-particle separation) and distribute their binding energy to nearby particles.

\subsubsection{Collisions During Binary Interactions}
\label{ss:collbinint}

{Scattering encounters between stars and binaries can enhance the rate of stellar collisions by orders of magnitude \citep[e.g.,][]{1987IAUS..125..187V}, largely due to close passages that can occur during chaotic resonant encounters \citep[e.g.,][]{1985ApJ...298..502H,1996ApJ...467..348M}. Furthermore, hydrodynamical simulations have suggested that when a collision does occur, the merger product can have a radius anywhere from 2 to 30 times larger than the sum of the two parent stars' radii \citep{2003MNRAS.345..762L}.  This increase in the collisional cross-section also increases the probability that the merger product will merge with another star \emph{during the same scattering encounter}  \cite[e.g.,][]{Fregeau2003}.  To account for this in \texttt{fewbody}, we modify the standard sticky sphere approximation as follows: whenever two stars merge, their merger product is assigned a mass equal to the sum of the two component masses, while the new radius is set to $f_{\rm exp}(R_1 + R_2)$, where $f_{\rm exp}$ is a constant factor (3, by default).  The encounter is then allowed to continue to determine whether the objects will merge again.  Whenever multiple collisions occur in a single encounter, we record the order of the collisions, and once the encounter is complete, the stars are then merged one at a time (in that same order) using the prescriptions in Section \ref{sec:sscoll}.}

{For BH-BH collisions, we set the collision radius to be 5 times the sum of the Schwarzschild radii of the two BHs ($10M$ in $G=c=1$ units, where $M$ is the sum of the two BH masses); this ensures that \texttt{fewbody} is operating in a regime where the pN expansion is valid.  For BBH mergers it is useful to know what the properties of the merging BHs were just prior to merger (e.g., eccentricity). Because the orbit of inspiraling BBHs is highly non-Newtonian at $10M$, we record the (Newtonian) semi-major axes and eccentricities of merging binaries at separations of 10M, 50M, 100M, and 500M.  This allows the orbit to be integrated forward in post-processing to determine the eccentricity of the binary at a given GW frequency.  Typically, the user should select the largest separation where the two BHs are bound, in order to minimize the error introduced by fitting to a Newtonian orbit \cite[see][Figure 2, bottom panel]{Rodriguez2018c}}.  

{Unlike the traditional sticky sphere approximation, the behavior of BBH merger products is modeled using detailed fitting formula from numerical relativity simulations, with the final mass, final spin, and recoil kick being calculated from the set of formula collected in \cite{Gerosa2016}.  When a merger occurs during a \texttt{fewbody} integration, the new BH mass and spin are set using these formulae, and the recoil kick is applied to the new merger product using the binary’s orbital angular momentum vector to determine the coordinate system for applying the kick.  This is useful for modeling the possibility of multiple BH mergers during a single scattering encounter, as well as the post-encounter orbital motion of any binaries in the cluster.  See \cite{Rodriguez2018c,Rodriguez2018} for details.  Note that these prescriptions for BH mass, spin, and recoil kick are also applied to any BBHs that merge inside the cluster.}

\subsection{Binary Formation}
\label{sec:binaries}

While star clusters are likely born with a non-negligible fraction of their stars in binaries \cite[e.g.,][]{Ivanova2005,Milone2012}, they can also dynamically produce binaries many Myr after their initial phase of star formation is complete.   Modeling this physics correctly is critical to understanding the long-term evolution of DSCs.

\subsubsection{Three-body Binary Formation} \label{threebodybinaries}

Two-body encounters cannot create bound binaries from single stars (at least for Newtonian point masses) without some way to dissipate energy from the encounter.  But if a two-body encounter occurs in the presence of a third body, excess kinetic energy can be carried away by the interloping particle, leaving behind a gravitationally-bound system.  While this process can involve more than three particles \cite[e.g.,][]{Tanikawa2013a}, \texttt{CMC} uses a probabilistic treatment of three-body binary formation \citep{Morscher2013} based on the formalism from \cite{Ivanova2005,Ivanova2010} and \cite{Oleary2006}.

Before either two-body relaxation or strong encounters are considered, the sorted list of stars is parsed three at a time, and the rate of binary formation for two objects of masses $m_1$ and $m_2$ from interactions with a third star ($m_3$) is calculated as

\begin{align}
\Gamma_{\rm 3bb} (\eta \geq \eta_{\rm min}) =& \sqrt{2} \pi^2 G^5 n^2 \left<w\right>^{-9}(m_i+m_{i+1})^5 \eta_{\rm min}^{-5.5} (1+2\eta_{\rm min}) \nonumber\\
& \times \left[1+2\eta_{\rm min}\left(\frac{m_i + m_{i+1} + m_{i+2}}{m_i + m_{i+1}}\right)\right]~,
\label{eqn:threebb}
\end{align}

\noindent where $n$ is the local stellar density of the cluster at that point, $\left<w\right>=4\sigma/\sqrt{3\pi}$ is the average relative velocity at infinity {for two particles in a Maxwellian distribution, and $\sigma$ is the local 3-D velocity dispersion}.  The rate is computed as a function of the binary hardness ratio, defined as

\begin{equation}
\eta \equiv \frac{G m_i m_{i+1}}{a \left<m\right> \sigma^2}~,
\end{equation}

\noindent where $a$ is the semi-major axis of the hypothetical binary, $\left<m\right>$ is the average local mass, and $\sigma$ is the average local velocity dispersion, both computed over the closest 20 particles to the encounter.  Note that this is half the number of particles used for computing the averages in Section \ref{sec:beta}, since three-body binary interactions are a fundamentally local process.    Since only hard binaries are expected to survive long enough to influence the long-term evolution of the cluster, we only form binaries with a minimum hardness of $\eta_{\rm min} = 5$.  By default, \texttt{CMC} only considers three-body binary formation for neighboring triplets of BHs in the cluster.  Both of these choices are available as user parameters in \texttt{CMC}. 

{We note that equation \eqref{eqn:threebb} differs from the original presentation in \cite{Ivanova2005,Ivanova2010} in two key ways.  In \cite{Ivanova2005} the rate of three-body binary formation depends on the distinct number densities of the three masses.  We simplify the expression here by replacing $n_2$, $n_3$, and $n_c$--the number densities of $m_2$, $m_3$, and of the overall cluster core--with $n$, the local number density of the cluster.  But for a given BH, we only consider $\Gamma_{\rm 3bb}$ whenever \emph{both} adjacent field stars are also BHs, meaning that in Monte Carlo fashion the probability calculated from equation \eqref{eqn:threebb} is implicitly multiplied by $(n_{\rm BH}/n)^2$ (or $n_{2} n_3 / n^2$ for particles with different masses); see also the discussion regarding binary-binary encounters in Section \ref{binint}.  In practice the majority of three-body binary formation occurs in the BH-dominated region of the core (even when three-body binaries are allowed to form from any stars), meaning that $n_c \approx n_2 \approx n_3 \approx n_{\rm BH}$ for most triple encounters.  }

{The second key difference concerns the relative velocities: when computing equation \eqref{eqn:threebb}, we use $\left<w\right>$, the averaged relative velocity between two particles.  But the original expression from \cite{Ivanova2005} also contains the terms $v_{12}/\left<w\right>$, where $v_{12}$ is the relative velocity between $m_1$ and $m_2$, and $v_{3}/\left<w\right>$, where $v_3$ is the relative velocity between $m_3$ and the $m_1$/$m_2$ center of mass. For computing the probability of binary formation, we assume $\left<w\right> = v_{12} = v_{3}$ for computational convenience.  However, when a binary is formed, we explicitly calculate $v_{12}$ and $v_3$ using the same coordinate system presented in Section \ref{sec:coord}.  This is necessary to determine the relevant energies of the encounter, and to sample the correct distribution of $\eta$ for that binary. We draw a random $\eta$ between $\eta_{\rm min}$ and $50$ using the normalized distribution given by $\frac{d\Gamma_{\rm 3bb}}{d\eta}$, and use this to determine the semi-major axis of the new binary (the eccentricities are assumed to be thermal).  The relative velocities of the binary's center-of-mass and the third particle are then adjusted to conserve energy.}

\subsubsection{Two-body Binary Formation}
\label{twobodybinaries}

While two-body binary formation is impossible when the two masses are Newtonian point particles, relaxing either of those assumptions can allow for binaries to form during close two-body encounters.  If at least one of the masses has a non-negligible radius (i.e., is not a white dwarf, neutron star or BH), close encounters can raise tides on one (or both) of the objects, allowing kinetic energy to be deposited into the structure of the stars themselves \cite[e.g.,][]{Fabian1975}.  If this energy exceeds the overall kinetic energy of the encounter, the two stars can form a binary through tidal capture. {We assume polytropic stellar models for the main-sequence stars undergoing tidal capture processes, and use polytropic index $n=1.5$ for low-mass main-sequence stars ($< 0.7\,\rm{M_{\odot}}$) and $n=3$ for higher-mass main-sequence stars. The calculations of the induced oscillation energy of the main-sequence stars and the cross sections during tidal capture follow prescriptions in \cite{PZwart_Meinen_1993} and \cite{Kim_Lee_1999}, respectively. We track the first passage of a single-single close encounter. If a star is tidally captured after this passage, we set the binary semi-major axis to be twice the pericenter distance immediately after tidal capture, assuming angular momentum conservation. Note that giant stars are not included for tidal capture because of the highly uncertain reaction of their envelopes to the tidal force. In addition to tidal capture, if a compact object or a main-sequence star collides with a giant star during single-single interactions, the compact object/main-sequence star can form a binary with the giant core by transferring the orbital energy to the giant envelope and ejecting it. We treat main-sequence stars as point mass during this process, and calculate the cross sections and final binary properties from the giant collisions using equations 4-6 in \cite{Ivanova+2006wd}.} See \cite{YeInPrep} for a complete description of these processes.

If instead of extended point particles we relax the Newtonian assumption, it becomes possible to form binaries through GW emission during close encounters (as occurs during BH mergers in \texttt{fewbody}).  In \texttt{CMC}, the maximum pericenter distance where the GW emission of the two BHs carries away sufficient energy to take the two particles from unbound (positive total energy) to bound (negative total energy) is calculated as \citep{Quinlan1987}:

\begin{equation}
r_p = \left(\frac{85\pi}{6}\right)^{2/7} \frac{G (m_{i}+m_{i+1})}{c^{10/7} w^{4/7}}~.
\end{equation}

BH binary formation via GW emission is decided probabilistically following a similar procedure to the collisions described in Section \ref{threebodybinaries}, using the above value of $r_p$ in equation \eqref{Eq:sigma_strong}.  When a binary is formed, its impact parameter at infinity is sampled uniformly within the area of its capture cross-section.  The semi-major axis and eccentricity are calculated from conservation of energy and angular momentum, accounting for the loss due to GW emission.  Many of these binaries are extremely wide (with semi-major axes of hundreds of AU or more), but their similarly extreme eccentricities $(e \gtrsim 0.99)$ mean they may still merge before they interact with neighboring particles.  For these BBHs, we compare the timescale for GW emission to drive the binary to merge. {We then compute the the rate for this binary to encounter other stars using  equations \eqref{eqn:rateofthings} and \eqref{eqn:rateofbs} (albeit with the actual mass and semi-major axis of this binary, instead of the averaged quantities), and in true Monte Carlo fashion, draw a random encounter time from an exponential distribution with that rate.  If the encounter occurs before the BBHs would merge and the binary is pathologically wide (as described in Section \ref{binint}) then the binary is disrupted.  Otherwise, its semi-major axis and eccentricity are evolved by integrating the orbit-averaged equations for $\frac{da}{dt}$ and $\frac{de}{dt}$ from \cite{Peters1964}.}

\subsection{Stellar Evolution}

\texttt{CMC} evolves every star and binary in the cluster forward by timestep $\Delta T$ using detailed prescriptions for stellar evolution.  These prescriptions, originally from \cite{Hurley2000,Hurley2002} have been significantly updated over many years, and form the basis for the \texttt{COSMIC} population synthesis package, which is now directly integrated into \texttt{CMC}.  See Section \ref{sec:cosmic}, \cite{Chatterjee2010}, and \cite{Breivik2020} for more details.  

{We perform stellar evolution in \texttt{CMC} after the dynamical encounters have taken place, but before the new cluster potential and orbits are computed.  This is to ensure that any dynamical changes to individual stars and binaries (particularly collisions and mass transfer in post-encounter binaries) can be accounted for using the same stellar physics, but before the new cluster orbits are calculated.  Collisions in \texttt{fewbody}, where multiple stars can potentially collide during a single encounter, are passed to the stellar evolution module in the same order that they occur after the encounter is complete; see also Section \ref{ss:collbinint}.}

\subsection{New Positions and Velocities} \label{S:orbitevolution}

At the end of every Monte Carlo timestep, a new position and corresponding velocity are chosen for each particle by randomly sampling a point on its orbit, weighted by the amount of time the particle spends at that point along its orbit. We do so by first computing the orbital pericenter $r_{\rm min}$ and apocenter $r_{\rm max}$ as the two roots of the energy equation:

\begin{equation}
Q(r) = 2 E - 2 \Phi(r) - J^{2}/r^2 = 0.
\label{eqn:energy}
\end{equation}

The probability $P(r)$ of finding the particle at a position $r$ within interval $dr$ is proportional to the time the particle spends traversing $dr$. Therefore,

\begin{equation} \label{Eq:orbit_probability}
P(r)dr = \frac{dt}{T} = \frac{dr/|v_r|}{\int_{r_{\rm min}}^{r_{\rm max}} dr/|v_r|},
\end{equation}
where $T$ is half the orbital period and $|v_r| = \sqrt{Q(r)}$. We draw a random sample from $P(r)$ and use the resultant $r_{\rm new}$ as the particle's position for the next timestep. Note that $P(r)\propto 1/|v_r|$ becomes infinite at the turning points of the orbit, where $|v_r|=0$. Changing coordinates by defining a suitable function $r=r(s)$ solves this problem \citep[see Section 2.6 of][]{Joshi2000}.  We must also set a new velocity consistent with the new orbital position: the new radial velocity is given magnitude $|v_r^{\rm new}| = \sqrt{Q(r_{\rm new})}$ with a randomly-selected sign, while the new transverse velocity is given by $V_t^{\rm new}=J/r_{\rm new}$.  Once the new positions and velocities have been calculated, the particles are then resorted in order of radial distance. 

\subsection{Escaping Particles}
After new positions and velocities have been sampled, we remove any particles that are no longer bound to the cluster.  For clusters in isolation, this is accomplished by removing any particles with positive energy after the dynamics and sorting steps are complete.  Such objects are routinely formed during strong encounters (which can eject particles from the central regions of the cluster) or through standard two-body relaxation, as many particles will slowly diffuse to positive energies.

\texttt{CMC} can also model clusters that are tidally limited, assuming the clusters to be on circular orbits within their galaxies.  For a cluster of mass $M_C$ on a circular orbit about the center of a point-mass galaxy, the tidal boundary of a cluster, $r_t$, can be calculated as

\begin{equation}
\label{eqn:tide}
r_t^3 = \left(\frac{M_C}{3M_G}\right)^{1/3} R_G,
\end{equation}

\noindent where $M_G$ is the mass of the galaxy and $R_G$ is the distance from the cluster to that galaxy \cite[e.g.,][]{Spitzer1987}.  Although sometimes called the Jacobi radius in stellar dynamics, equation \eqref{eqn:tide} is identical to the calculation of the Roche surface in binary stellar physics.  As the cluster loses mass, the initial tidal boundary (specified by the user) decreases at a rate $\propto M_C^{1/3}$, following equation \eqref{eqn:tide}.\footnote{While equation \eqref{eqn:tide} is only correct for a point-mass galactic potential, the $M_C^{1/3}$ scaling is correct for any circular orbit in a spherical potential \cite[e.g.,][]{renaud2011}, meaning this approach is always correct provided that the orbit remains circular and the correct initial $r_t$ is specified.}

Of course the Jacobi surface is not spherical; $r_t$ actually represents the location of the Lagrange points along the line from the cluster to the galactic center. Since the H\'enon method assumes the cluster to be spherically symmetric, approximate methods must be used to determine whether a star is stripped from the cluster by the galactic potential.  \texttt{CMC} uses an energy-based criterion \citep{giersz2008} that removes any star with an energy above the potential at the tidal radius

\begin{equation}
\label{eqn:tideneregy}
E > \alpha \Phi(r_t),
\end{equation}

\noindent where the parameter $\alpha$ is given by

\begin{equation}
\alpha = 1.5 - 3\left(\frac{\ln \Lambda}{N}\right)^{1/4}.
\end{equation}

\noindent As previously, $N$ is the initial number of stars in the cluster and $\ln \Lambda$ is the Coulomb logarithm from \eqref{eqn:loglambda}. Numerical testing \citep{giersz2008,Chatterjee2010} has shown that equation \eqref{eqn:tideneregy} produces significantly better agreement to direct $N$-body simulations than simply stripping stars with apocenters beyond the tidal boundary.   

\subsection{Potential Calculation} \label{S:potential}

Because we assume the cluster potential to be spherically symmetric, computing it is straightforward.  The potential at point $r$, located between two particles at $r_i$ and $r_{i+1}$, is given by

\begin{equation} \label{Eq:potential1}
\Phi(r)=-G\left(\frac{M_i}{r} + \sum_{i=j+1}^N\frac{m_i}{r_i}\right),
\end{equation}
where $M_i\equiv\sum_{j=1}^i m_j$ is the total mass interior to star $i$.

While equation \eqref{Eq:potential1} is formally correct, it is computationally expensive to evaluate whenever the potential at an arbitrary radius is required (such as when computing new orbital positions).  However, we can accelerate this process by evaluating the potential at every particle, $\Phi_i$, and interpolating the potential between that particle and its nearest neighbor, $\Phi_{i+1}$.  Because the stars are already sorted by radius, the potential at each particle can be computed recursively from the outermost star inwards with

\begin{equation} \label{Eq:potential2}
\Phi_{N+1}=0,\ \ \ \ \Phi_i=\Phi_{i+1}-G M_i \left(\frac{1}{r_i}-\frac{1}{r_{i+1}}\right),
\end{equation}

\noindent where $M_N = \sum_{i=1}^N m_i$ is the total mass, and $M_{i-1}=M_i-m_i$.  With $\Phi_i$ evaluated for every particle in the cluster, the potential at any radius  $r_i < r < r_{i+1}$ can be expressed as

\begin{equation} \label{Eq:potential3}
\Phi(r)=\Phi_i + \left( \frac{1/r_i-1/r}{1/r_i-1/r_{i+1}}
\right)\left(\Phi_{i+1}-\Phi_i\right).
\end{equation}

The potential at any $r$ is then computed by finding the index $i$ such that $r_i\leq r\leq r_{i+1}$ and applying equation \eqref{Eq:potential2}.

\subsection{Global Timestep Selection} \label{S:timestep}

In \texttt{CMC}, the timestep $\Delta T$  is computed as the minimum of several important physical timescales. The first such timescale is the local average relaxation timescale -- the average time to deflect a particle by angle $\beta_{e}^{\rm max}$ within a certain localized region of the cluster.  Following equation \eqref{Eq:relax7}, we define this as

\begin{equation}
\widehat{T}_{{\rm{rel}},i} = \left(\frac{\beta_{e}^{\rm max}}{\pi/2}\right)^2 \frac{\pi}{32} \frac{\left< w\right>^3}{4 G^2 n \langle m^2 \rangle \ln \Lambda},
\label{eqn:timesteprelax}
\end{equation} 

\noindent where $\beta_{e}^{\rm max}$ can be set by the user, with larger $\beta_{e}^{\rm max}$ corresponding to larger relaxation timesteps.  By default, \texttt{CMC} sets $\beta_{e}^{\rm max} = \sqrt{2}$ (though we typically use $\beta_{e}^{\rm max}=1$ for systems of point masses). 
Whereas $T_{{\rm rel},i}$ is defined for pairs of particles $i$ and $i+1$, $\widehat{T}_{{\rm{rel}},i}$ is a broader rolling average over particles centered on particle $i$.  The local number density $n$ is calculated using the same rolling average as in equation \eqref{Eq:relax6}, with similar rolling averages used to compute $\left< m^2 \right>$ and the relative velocity between two particles, $\left< w \right>$, (calculated from the averaged 3-D velocity dispersion, assuming a locally Maxwellian velocity distribution).  To accurately capture the overall relaxation dynamics, we require that $\Delta T < \widehat{T}_{{\rm{rel},i}}$ for all $i$ particles.  In \citet[][equation 11]{Freitag2001}, this was accomplished by selecting a timestep in each radial bin that was smaller than $\widehat{T}_{{\rm{rel}},i}$ by a factor of 0.005 to 0.05.  In \texttt{CMC}, we instead choose a single relaxation timestep for the entire clsuter, taken to be the minimum $\widehat{T}_{{\rm{rel}},i}$ for all particles $i$. This choice ensures that our global timestep is sufficiently short for all particles and to resolve the complicated dynamics in the cluster core, while making the parallelization of the code significantly easier; see \cite{Pattabiraman2013}.

In addition to the relaxation timescale, we also compute characteristic timescales for collisions, ($T_{\rm{coll}}$) binary--binary encounters ($T_{\rm{bb}}$) and binary--single encounters, ($T_{\rm{bs}}$).  In general, the rate of collisions can be found by integrating the cross-section for strong encounters over the distribution function of velocities for the two particles \citep{Freitag2002,Binney2008}:

\begin{equation}
\Gamma_{\rm strong} = \frac{1}{n} \int F(\mathbf{v}_{i})F(\mathbf{v}_{i+1}) w\Sigma_{\rm strong}   d^3\mathbf{v}_i d^3\mathbf{v}_{i+1}~,
\label{argh}
\end{equation}

\noindent where as before $F$ is the distribution function (ignoring the mass and position), $w=\big| \mathbf{v}_{i+1} - \mathbf{v}_i \big|$ is the relative velocity, and $\Sigma_{\rm strong}$ is taken from equation \eqref{Eq:sigma_strong}.  Assuming a Maxwellian distribution for the velocity distribution, equation \eqref{argh} can be evaluated as \cite[e.g.][p.626]{Binney2008}

\begin{equation}
\Gamma_{\rm strong} = 4 \sqrt{\pi} n r_p^2 \sigma \left(1+\frac{G M}{2 r_p \sigma^2}\right)
\label{eqn:rateofthings}
\end{equation}

\noindent and the relevant timescale between encounters is simply $T_{\rm strong} = \Gamma_{\rm strong}^{-1}$.  The timescales of relevance to \texttt{CMC} are then \cite[see][for details]{Fregeau2007}

\begin{align}
    T_{\rm{coll}}^{-1} &= 16\sqrt{\pi} n_{s} \left< R^2 \right> \sigma \left(1 + \frac{G\left< MR\right>}{2\sigma^2 \left< R^2 \right>} \right),\\
    T_{\rm{bs}}^{-1} &= 4\sqrt{\pi} n_{s} X_{\rm{bin}}^2 \langle a^2 \rangle \sigma \left(1 + \frac{G\langle M \rangle \langle a \rangle}{2\sigma^2 X_{\rm{bin}}\langle a^2 \rangle} \right),\label{eqn:rateofbs}\\
    T_{\rm{bb}}^{-1} &= 16\sqrt{\pi} n_{\rm{bin}} X_{\rm{bin}}^2 \left< a^2 \right> \sigma \left(1 + \frac{G\left< M a\right> }{2\sigma^2 X_{\rm{bin}}\left< a^2 \right>} \right),
\end{align}

\noindent where all of the averaged quantities above, including the number densities of single and binary stars, $n_s$ and $n_{\rm bin}$, and the velocity dispersion $\sigma$, are computed over the innermost $300$ particles in the cluster (i.e.~the central regions of the core).  {Note that if no binaries are present within these 300 stars, $T_{\rm bs}$ and $T_{\rm bb}$ are set to infinity for that timestep.  Conversely, if \emph{only} binaries are present, then $T_{\rm bb}$ is computed normally, while $T_{\rm bs}$ is set to infinity.}

Finally, while the stellar evolution package has its own internal timestep, the mass loss from stars can drive significant dynamical changes in the cluster, particularly in the early stages of evolution.  We compute a relevant timescale for stellar evolution based on the total fraction of mass lost in the previous timestep: 

\begin{equation}
T_{\rm{se}} = 0.001  \left( \frac{M_{\rm{cl}}}{\Delta M_{\rm{se}}} \right)\Delta T_{\rm{prev}},
\label{eqn:timestepse}
\end{equation}

\noindent where $M_{\rm{cl}}$ is the total cluster mass, $\Delta T_{\rm{prev}}$ is the previous timestep and $\Delta M_{\rm{se}}$ is the mass lost during the current timestep.  \texttt{CMC} uses the minimum individual timestep from equations (\ref{eqn:timesteprelax}-\ref{eqn:timestepse}) as the global timestep for the cluster. {However, we note that for a typical cluster model (such as those presented in Section \ref{s:king}), the central relaxation timescale is the smallest of the relevant timescales for more than 96\% of the timesteps within the first 100 Myr.}

\subsection{Energy Conservation}
\label{sec:energyconv}

Finally, to ensure energy conservation, a correction must be made to the energies of the stellar orbits.  While the two-body relaxation of the H\'enon method intrinsically conserves energy, the process of calculating new orbits and sampling positions and velocities is done \emph{after} dynamical encounters have altered the energy and angular momenta of each orbit, but \emph{before} the potential has been updated to account for the changes in the orbits.  In other words, the $\Phi$ in equation \eqref{Eq:potential1} is perpetually one timestep behind the new $E$ and $J$, allowing this time-dependent shift in the potential to do work on the particle.  While the energy error during a single timestep is negligible, this effect can become significant during long-term integrations leading to a spurious drift in the energy of the cluster.  

To counter this, we adopt the energy conservation scheme first developed by \cite{Stodoikiewicz1982} in his update of the method.  First, note that the change in energy for a single particle is expressible as

\begin{equation}
\Delta E_i^{\rm corr} = \int \frac{\partial U(r_i)}{\partial t} dt~.
\label{eqn:energyconv}
\end{equation}

\noindent Since we are interested in the work done by the change in potential between the previous and current timesteps, equation~\eqref{eqn:energyconv} can be approximated as the average of the change in potential energy between the positions of the particle at the previous and current timesteps:

\begin{equation}
\int \frac{\partial U(r_i)}{\partial t} dt = \left[\Delta \Phi(r_i^{\rm prev})+\Delta\Phi(r_i^{\rm curr})\right]/2~,
\end{equation}

\noindent where $\Delta\Phi \equiv \Phi^{\rm curr} - \Phi^{\rm prev}$.  To implement this conservation scheme, \texttt{CMC} records the four relevant energies for each particle every timestep --- $\Phi^{\rm prev}(r_i^{\rm prev})$, $\Phi^{\rm prev}(r_i^{\rm curr})$, $\Phi^{\rm curr}(r_i^{\rm prev})$, and $\Phi^{\rm curr}(r_i^{\rm curr})$ --- and uses them to compute $\Delta E^{\rm corr}_{i}$ for each particle.  The velocity magnitudes for each particle are then scaled to increase or decrease the particle's kinetic energy by $\Delta E_i^{\rm corr}$ (while maintaining the ratio of transverse to radial velocities as determined by the H\'enon method), ensuring that the cluster conserves energy over many relaxation times.  

In addition to the energy drift, a second issue arises from calculating stellar orbits for the current timestep using the potential from the previous timestep.  For any given particle, the potential includes the contribution \emph{of that same particle} at its last position in the cluster, and when the new orbit is calculated, it includes this spurious ``self-force'' from the previous timestep.  While this effect is relatively minor for clusters with equal-mass particles, it can be significant in clusters with a realistic mass spectrum, particularly when the most-massive particles (e.g.~BHs) are concentrated in the cluster center where only a handful of particles determine the potential \citep{Freitag2001,Fregeau2007}.

To solve this, we add a correction term to remove the particle's own contribution to the previous potential at its previous position:

\begin{equation}
\Phi_{\rm sf}(r) = \begin{cases}
G M_i/r&~~~r\geq r_i^{\rm prev},\\
G M_i/r_i^{\rm prev}&~~~r<r_i^{\rm prev}.
\end{cases}
\end{equation}

\noindent This self-force correction, $\Phi_{\rm sf}(r)$, is added to the energy equation \eqref{eqn:energy} when the new radial positions are sampled from the particle orbits (Section \ref{S:orbitevolution}).  

\subsection{Limitations of the Method}
\label{S:caveats} 

The Monte Carlo method, and H\'enon's method in particular, has been used for 50 years for the study of dense star clusters, starting with the initial papers by \cite{Henon1971a,Henon1971b}.  The technique was reintroduced with improvements by \cite{Stodoikiewicz1982}, and it is this version that most modern implementations \citep{Giersz1998,Joshi2000,Freitag2001,2014MNRAS.443.3513S} are based upon.  While Monte Carlo methods have enjoyed great success in the modeling of dense star clusters, that success relies on a series of assumptions which we have implicitly or explicitly made in Section \ref{S:code}.  We now briefly describe each of these assumptions and the limitations they impose upon the method.  

Underlying nearly all of the formal averages and encounter rates presented in Section \ref{S:code} is Boltzmann's \textbf{molecular chaos assumption}, where we have assumed that the distribution functions of our test stars and field stars are independent and separable.  This assumption is explicitly used in equations \eqref{argh2} and \eqref{argh}, where we assume that the joint distribution function for particles $i$ and $i+1$ can be written as 

\begin{equation}
\label{eqn:seperabolity}
F(\mathbf{v}_{i},\mathbf{v}_{i+1}) = F(\mathbf{v}_{i})F(\mathbf{v}_{i+1})~.
\end{equation}

\noindent Equation \eqref{eqn:seperabolity} can be justified in the following way: since equal octaves of impact parameter contribute equally to equation \eqref{eq:int}, the majority of two-body encounters will occur at $b_{90} \ll b \ll b_{\rm max}$. Because of this, it is highly unlikely for two particles in a sufficiently large cluster to encounter each other more than once within their lifetimes, or at least before other encounters have ``reset'' the velocities of the two particles.  While this assumption is valid for large-$N$ systems, it does mean that the Monte Carlo method cannot resolve resonant effects between particles, such as scalar resonant relaxation \citep{1996NewA....1..149R} around central-massive BHs or resonant behavior between heavy objects in the centers of GCs \citep{2019ApJ...878..138M}.

The assumption of large $N$ is also employed when computing the two-body diffusion of $(\Delta v)^2$.  When computing equation \eqref{Eq:relax35}, we assumed $\Lambda = b_{\rm max}/b_{90} \gg 1$, which allowed us to substitute $\ln \left(1 + \Lambda^2\right)\simeq 2\ln(\Lambda)$, and neglect any higher-order contributions to the change in velocity e.g., $\left<(\Delta v_i)^4\right>$.  This neglect of the higher-order terms is known as the \textbf{Fokker-Planck approximation}, and is largely justified because any terms beyond second order are at least a factor of $\ln(\Lambda)^{-1}$ smaller than the $\left<(\Delta v)^2\right>$ term \citep{1973dses.conf..183H,Binney2008}.  This is also related to our assumption of \textbf{orbit averaging}, where we assumed $N$ is sufficiently large that $T_{\rm relax} \gg T_{\rm dyn}$ in equation \eqref{eqn:approxtrel}, allowing us to treat the orbits as essentially fixed on dynamical timescales.  

Naturally one may ask how large must $N$ be for the above assumptions to be valid.  While hard to derive explicitly,

\begin{equation}
N \gtrsim 10^3 \frac{m_{\rm max}}{\left<m\right>}
\end{equation}
 
\noindent has been shown to be a decent working criterion \citep{Freitag2008}.  Note that this criterion depends on the ratio between the largest mass in the cluster, $m_{\rm max}$, and the average stellar mass, $\left< m \right>$.  For clusters with realistic initial mass functions, the ratio can be as large as 100:1 (once stellar evolution has reduced the mass of the most massive stars). In our experience, $N$ must be $\gtrsim 10^5$ initially for such clusters to show acceptable energy conservation and good agreement with direct $N$-body results  \citep{Chatterjee2010}.

H\'enon's method also assumes that the cluster is \textbf{spherically symmetric}.  This assumption is not explicitly related to the size of the cluster or the relaxation timescale, but is necessary for computing new particle orbits, determining nearest neighbors, and computing the potential.  While this is an appropriate assumption for many GCs and some NSCs, this does mean that clusters with net angular momentum (e.g., rotating GCs, if the rotation creates a significant departure from spherical symmetry) or triaxial stellar distributions (e.g., NSCs that have recently undergone a merger) cannot be modeled with \texttt{CMC}.  To address this, similar techniques, referred to as orbit-following Monte Carlo methods, have been developed that explicitly follow the orbits of stars in arbitrary potentials on a dynamical timescale, either by directly integrating the orbits in a given potential \citep[the so-called Princeton Monte Carlo, of][]{1971ApJ...164..399S,1972ApJ...175...31S} or by tracking the diffusion of orbits over an integer number of orbital periods \citep[the so-called Cornell Monte Carlo of][]{1980ApJ...239..685M}.  These techniques and have been used extensively for the study of triaxial NSCs and central-massive BH dynamics \cite[see e.g.,][for recent examples]{Hopman2009,Madigan2011,Vasiliev2014}.  We note that the assumption of spherical symmetry does \emph{not} mean that we have assumed the velocities to be isotropic. \texttt{CMC} can model systems with significant velocity anisotropy, since the ratio of $v_r$ to $v_t$ is calculated self-consistently every timestep.  

Finally, because the H\'enon method operates on a relaxation timescale, we are implicitly assuming that GCs can be modeled as quasi-adiabatic transitions between different cluster models in \textbf{dynamical (virial) equilibrium}.  While this assumption is largely valid for the long-term evolution of clusters in most cases, there are notable exceptions, such as tidal shocking \citep{1972ApJ...176L..51O}, or the initial ``violent relaxation'' of the cluster after its (potentially highly non-virialized) birth \citep{1967MNRAS.136..101L}.  These processes involve clusters that are sometimes far from equilibrium and evolve to a virialized state on a dynamical timescale.  For these cases, more computationally expensive orbit-following techniques, such as the aforementioned Princeton/Cornell methods or direct $N$-body integrations, must be employed.

\section{{\tt CMC} Package Overview} \label{S:release}
\label{S:cmc-cosmic}
The public release of \texttt{CMC} coincides with the latest release of the \texttt{COSMIC} package for binary population synthesis \citep{Breivik2020}, which is included directly in the \texttt{CMC} GitHub repository (and can be compiled simultaneously).  Both codes implement single stellar evolution using \texttt{SSE} \citep{Pols1998,Hurley2000} and binary evolution using \texttt{BSE} \citep{Hurley2002}, with many major updates having been made to the physical prescriptions of stellar and binary evolution based on advances over the past two decades; see \cite{Breivik2020} for details. 

With the inclusion of \texttt{COSMIC} into \texttt{CMC}, we have also updated both codes to use similar input/output files and initial condition generators.  Both codes use identical parameters and initialization files, allowing cluster simulations and population synthesis to be easily performed with identical physics.  The generation of all initial conditions for both binary populations and cluster simulations is now included in the latest version of \texttt{COSMIC}, allowing populations with identical physics to be created from a simple Python interface.

\subsection{Updates to \texttt{COSMIC} Population Synthesis Code}
\label{sec:cosmic}

While \texttt{COSMIC} was originally based on the version of \texttt{BSE} incorporated into \texttt{CMC}, including its updates to compact object physics \citep{Chatterjee2010,Rodriguez2016}, the two codes have diverged over the past several years, and the use of \texttt{COSMIC} as a community-driven population synthesis code has kept it up-to-date with the latest developments in binary stellar evolution.  Here we review the changes to binary stellar evolution (and in particular binary mass transfer) in the latest version of \texttt{COSMIC} (v3.4), which is paired with the release of \texttt{CMC}.  For a complete description of the physics included in \texttt{COSMIC} and the various options, see \cite{Hurley2000,Hurley2002,Breivik2020}

Since the release of \texttt{COSMIC} v3.3, the treatment of Roche-overflow mass transfer from non-degenerate stars has been expanded to include multiple assumptions for both mass loss rates from the donor and mass accretion rates onto the accretor. The default mass loss rate is determined using the expression given by \citet{Hurley2002}, which steeply increases with the ratio of the donor radius to its Roche lobe radius 

\begin{equation}
    \label{eq:hurley_don_lim}
    \dot{M}_{\rm{don}} = F(M_{\rm{don}})\left[\ln\frac{{R_{\rm{don}}}}{R_{\rm{rl},\rm{don}}}\right]^3 \, {M}_{\odot}\, \rm{yr}^{-1}, 
\end{equation}
\noindent where $\dot{M}_{\rm{don}}$ is the donor mass loss rate, $M_{\rm{don}}$ is the donor mass in units of $M_\odot$, $R_{\rm{don}}$ is the donor radius, $R_{\rm{rl},\rm{don}}$ is the donor's Roche lobe radius, and 
\begin{equation}
    F(M_{\rm{don}}) = 3\times10^{-6}\, [\min \left( M_{\rm don}, 5.0\right)]^2~.
\end{equation}

Version 3.4 of \texttt{COSMIC} also includes the Roche-lobe overflow prescription of \citet{Claeys2014}, which is calibrated to binary systems with Zero Age Main Sequence component masses less than $10\,M_{\odot}$.
In this case, the mass loss rate of the donor is determined similarly to equation~\eqref{eq:hurley_don_lim}, but with an overall multiplicative factor, $f$, given by

\begin{equation}
    \label{eq:claeys_don_lim}
    f = \left\{
        \begin{array}{ll}
            1000, & \quad Q < 1 \\
            \rm{max}\big(1, \frac{1000}{Q}\exp \big[-\frac{1}{2} \big(\frac{\ln Q}{0.15} \big)^2 \big]\big) & \quad Q \geq 1,
        \end{array}
    \right.
\end{equation}
\noindent where $Q \equiv M_{\rm{acc}} / M_{\rm{don}}$. In both cases, mass loss is always limited by the thermal timescale such that
\begin{equation}
    \label{eq:thermal_lim}
    \dot{M}_{\rm{don}}^{\rm{max}} = \max \left(\dot{M}_{\rm don }, \frac{M_{\rm{don}}}{\tau_{\rm{KH}}} {M}_{\odot} \rm{yr}^{-1}\right),
\end{equation}
\noindent where

\begin{equation}
    \label{eq:tkh}
    \tau_{\rm{KH}} = 3.1\times 10^7 \left(\frac{M}{{M}_{{\odot}}}\right)^2\left(\frac{{R}_{\odot}}{R}\right)\left(\frac{{L}_{\odot}}{L}\right) \rm{yr},
\end{equation}

\noindent is the Kelvin-Helmholtz time.

Four prescriptions are now included to determine the amount of mass that the accretor is allowed to gain, in addition to the original one in  \citet{Hurley2002}. The original \texttt{BSE} prescription assumes that stars with radiative envelopes (i.e., main sequence, Hertzsprung gap, and core helium burning) can accrete mass at a rate limited to ten times the thermal rate of the star, while stars with convective envelopes (i.e., giant branch, asymptotic giant branch) can accrete mass at an unlimited rate. 
Three of the newly added prescriptions consider modifications to this prescription: one that limits stars with radiative envelopes to accrete mass at the thermal rate, one that applies accretion limits of ten times the thermal rate for all stars---not just those with radiative envelopes---and one that applies accretion limits of the thermal rate for all stars. We also include another choice where accretion is a fixed fraction of the mass lost by the donor, as used in \cite{Belczynski2008}.

\subsection{Cluster Initial Conditions} 

In \texttt{COSMIC}, initial conditions are generated by sampling particle positions and velocities from either a \cite{Plummer1911}, \cite{Elson1987}, or \cite{King1966} profile.  Both the Elson and Plummer profiles are sampled piecewise, with the particle positions drawn first from the cumulative mass distribution for each given profile.  To self-consistently sample velocities for each particle at its given position, we draw velocities from the distribution function via rejection sampling up to the local escape speed \citep{Aarseth2003}.  For the Plummer profile we use the analytic form of the distribution function, while for the Elson profile we numerically solve for the distribution function in energy space \cite[see e.g.,][Appendix B]{grudic2018}.\footnote{We note that previous $N$-body samplers of the Elson profile, such as the one described in \cite{2011MNRAS.417.2300K}, have used the 1-D Jeans equations to sample the velocity dispersion of the cluster.  While this yields a cluster in virial equilibrium, it does not produce a cluster consistent with a single distribution function.}  The King profile is generated in a similar fashion: the differential equations for cluster density, potential, and enclosed mass are solved numerically, with the latter being used to sample random positions from the cluster center.  The velocities are then sampled from the analytic distribution function via the aforementioned rejection sampling technique.

By default, \texttt{COSMIC} draws radii and velocities from the available distributions before converting them into \textit{H\'enon} units \citep{Heggie1986,Heggie2014}, where the gravitational constant and initial cluster mass are both set to unity ($G=M_0=1$) and {the sum of the initial cluster kinetic and potential energies is $E_{\rm 0,kin} + E_{\rm 0,pot} = - 1/4$. (This is not the same as the total initial cluster energy $E_0 = E_{\rm 0,kin} + E_{\rm 0,pot} + E_{\rm 0,bin}$, where $E_{\rm 0,bin}$ is the initial negative-valued total binding energy in primordial binaries, if any.)} The units of mass, length, and time are then

\begin{align}
&U_m = M_0 ~,\nonumber\\
&U_l = \frac{-GM_0^2}{4 E_0} ~,\nonumber\\
&U_t = \frac{G M_0^{5/2}}{(-4 E_0)^{3/2}}~,\label{eqn:unit}
\end{align}

\noindent respectively.  The above length units are such that the initial virial radius of the cluster is unity,  while the units of time roughly correspond to the crossing time of a particle at the virial radius.  However, because \texttt{CMC} operates on a relaxation timescale, we instead use a more appropriate unit of time,

\begin{equation}
U_t^{\rm rel} \equiv \frac{N_0}{\log \gamma N_0} U_t~,
\end{equation}

\noindent where $N_0$ is the initial number of particles (single stars or binaries).  These units are the default used throughout \texttt{CMC}.

Once the positions and velocities have been selected, the user can optionally add additional stellar physics to each of the particles, such as stellar masses, radii, and binary companions.  All of the options available in \texttt{COSMIC}, including initial mass functions (e.g.,~\citealt{Kroupa2001} or \citealt{Salpeter1955}), binary initial conditions (e.g.~following \citealt{Sana2012} or \citealt{Moe2017}), and stellar metallicities, can be used to generate realistic cluster initial conditions for \texttt{CMC}.   When generating binary initial conditions for cluster simulations, we truncate the upper limit of the $P_{\rm orb}$ distribution at the hard-soft boundary for each binary at its respective radius in the cluster \cite[see e.g.,][]{Heggie1975}.

\subsection{Input/Output}

  \begin{figure*}
\centering
\includegraphics[]{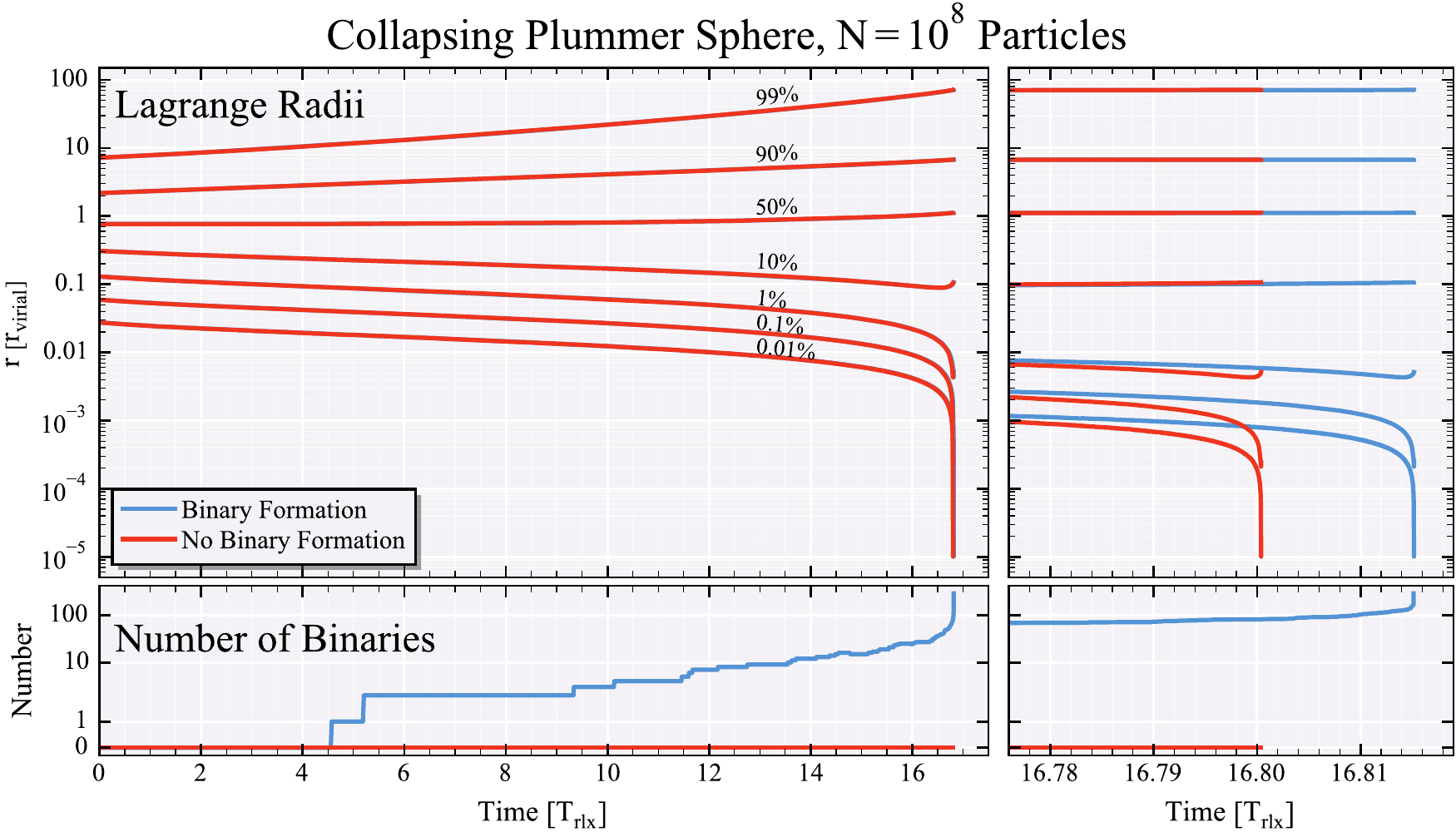}
\caption{The Lagrange radii (radius enclosing a given fraction of the mass) and number of binaries as a function of the initial half-mass relaxation time for the two $N=10^8$ particle realizations of a Plummer sphere.  We show in red the model without three-body binary physics, and in blue the same model, but allowing binaries to form from single stars.  On the \textbf{top}, we show the Lagrange radii as a function of time, with the distinction between the core and halo rapidly developed due to two-body relaxation.  On the \textbf{top right}, we zoom in at the final moments of core collapse (defined as having less than 1000 objects in the core), and clearly show the (minor) delay in core collapse caused by binary heating in the cluster core.  On the \textbf{bottom}, we show the number of binaries formed over time.  As the core shrinks, the central encounter rate increases, dramatically increasing the number of binaries in the cluster in the final moments before collapse.  }
\label{fig:lagrad}
\end{figure*}

Input in \texttt{CMC} is handled at the command-line level by passing an initialization file containing all the parameters controlling the physics, parallelization, and runtime options (unspecified parameters are set to default values).  The parameter files contain many of the same options (particularly related to stellar evolution) as the parameter files for \texttt{COSMIC}, allowing initial conditions, population synthesis, and cluster dynamics simulations to be performed with identical options.  The initial conditions are saved as 
tables in the Hierarchical Data Format (HDF5) and passed as input to the initial parameters file. 

Three forms of output are generated during a typical \texttt{CMC} run.  First, a series of tab-separated data files containing specific parameters of the cluster (e.g., mass, half-mass radius, number of binaries) are printed every timestep and can be easily read in any scripting language.  Second, specific events, such as stellar collisions, BH mergers, binary--single encounters, etc., are recorded as they occur in human-readable log files (with a provided Python parser).  Third, snapshots of the full cluster state are periodically written to several HDF5 tables at user-specified intervals.  Separate files exist for snapshots containing all stars in the cluster and those containing only the BHs (allowing the latter to be resolved at shorter intervals).   Each snapshot is saved in its respective file as a pandas-readable HDF5 table with keys corresponding to the simulation time when the snapshot was written.  These can be directly imported into Python using the \texttt{read\_hdf} command in pandas \citep{mckinney-proc-scipy-2010,jeff_reback_2021_4681666}.

Finally, \texttt{CMC} periodically saves checkpoints by directly dumping the state of every processor to a binary file associated with that processor.  These checkpoints allow bit-for-bit restarting of a \texttt{CMC} run should the run either be interrupted or terminated due to computing cluster queue-time limits.  While the checkpoints contain the state of the cluster, stars, and binaries up to that point, both the physical parameters of the run (set in the initialization file) and the current random number seed can be changed by the user at restart.

\subsection{Parallelization} \label{S:parallelization}
\texttt{CMC} is designed to be compiled against the standard libraries of the Message Passing Interface \cite[MPI;][]{mpi1994}, enabling distributed-memory parallelization across multiple cores and machines.  Many of the physical processes in \texttt{CMC}, such as stellar evolution and nearest-neighbor encounters, are fundamentally local processes that are amenable to simple parallelization.  However, the radial sorting of stars, the computation of new stellar orbits in the global cluster potential  (Section \ref{S:orbitevolution}),  and even the computation of the potential itself (Section \ref{S:potential}) are global processes that require a customized parallel approach.  

Because the calculation of the cluster potential in equation \eqref{Eq:potential3} and the nearest-neighbor dynamics require the particles to be sorted radially, the particles are resorted each timestep after new positions are selected (Section \ref{S:orbitevolution}) in order of increasing radius from the cluster center.  \texttt{CMC} uses a custom parallel sorting implementation, in which the local data on each processor is sorted using a \texttt{Quicksort} algorithm  \citep{Hoare1961} before being combined globally across processors using a parallel \texttt{Sample Sort} algorithm \citep{FrazerMcKellar1970}.  This implementation allows for rapid sorting and redistribution of particles across multiple distributed memory processes. For more details on the parallel sorting method, see Section 3.4 of \citet{Pattabiraman2013}.  Once the sorting is complete, the particles are redistributed across all MPI processes with a series of all-to-all communications.  

The calculation of the stellar orbits is more straight-forward, and only requires that every MPI process have access to the global potential of the cluster at each timestep.  To that end, each process maintains a global list of the radial positions and masses for every star in the cluster, allowing each process to compute equation \eqref{Eq:potential3} independently.  These lists are updated every timestep.  For a complete description of this and the other parallelization strategies employed in \texttt{CMC}, see \cite{Pattabiraman2013}.  

\section{Examples}
\label{S:examples}

Having described the code in detail, we now show two standard examples of clusters generated with \texttt{CMC}: a Plummer sphere of point masses evolved from its initial state to core collapse, and the evolution of 10 realistic King models of GCs evolved with all available physics for $12\,\gyr$.

\subsection{Plummer Sphere to Core Collapse}

  \begin{figure}
\centering
\includegraphics[]{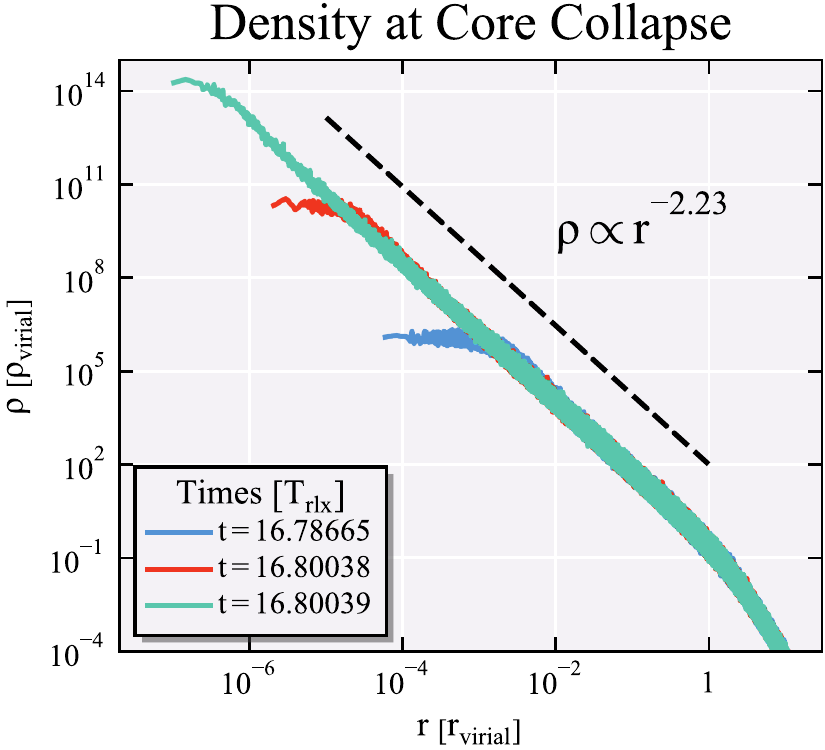}
\caption{The density profile of the Plummer sphere with $10^8$ particles near core collapse as a function of radius.  Both quantities are calculated in terms of the initial virial radius and the initial density of objects within that radius.  We show the densities from three separate snapshots, whose time is given in terms of the initial half-mass relaxation time.  During the last moments of the cluster's evolution, the inner regions rapidly converge to the $\rho \propto r^{-2.23}$ power-law expected from Fokker-Planck calculations \citep{Cohn1980}, showing self-similar behavior over more than 15 orders of magnitude in density.}
\label{fig:rhor}
\end{figure}

The Plummer profile was one of the first used to fit realistic observations of GCs to a theoretical model, and remains popular since, unlike models based on more sophisticated distribution functions, most of its key features can be expressed analytically \citep{Heggie2003}.  This also makes it a favorite tool of the $N$-body simulator since the enclosed mass, velocity dispersion, and distribution function can all be expressed analytically, and are easily sampled to generate cluster initial conditions.  Here, we use \texttt{COSMIC} to generate a realization of a Plummer sphere with $10^8$ initial point-mass particles.  We choose this particular setup because the evolution of a Plummer sphere to core collapse is a well-studied problem in the literature \citep[e.g.,][]{Freitag2001,Freitag2006b,Binney2008}. 
We then integrate these initial conditions twice using \texttt{CMC}: once with no binary physics, and once with three-body binary formation and strong binary encounters enabled.

  \begin{figure}[]
\centering
\includegraphics[]{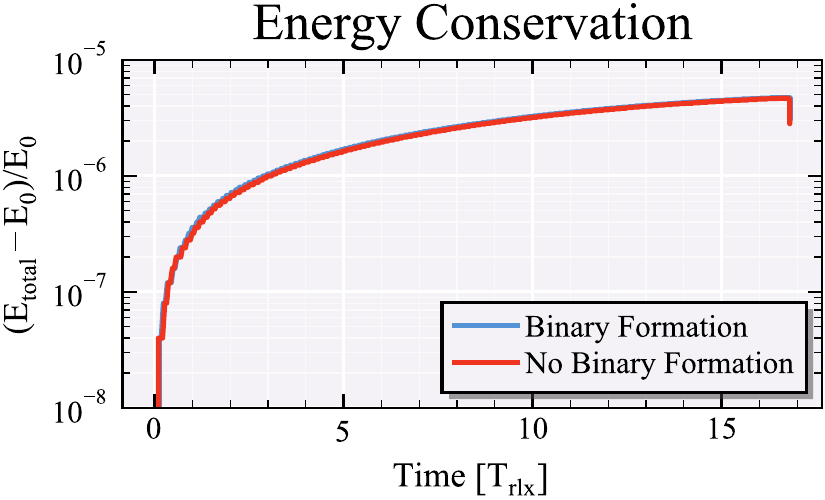}
\caption{The total energy of the two Plummer sphere models (including the energy of ejected particles) over time.  The sharp downturn at the end arises from the reduction of the timestep near core collapse to better resolve the dense central regions of the cluster.}
\label{fig:energy}
\end{figure}

In Figure \ref{fig:lagrad}, we show the evolution of the two clusters from their initial state to core collapse, defined as having $< 1000$ stars within the theoretical core radius \cite[see][for the description of the latter]{Casertano1985}.  We also show the number of binaries as a function of time, since it is the formation and hardening of binaries in the core that provides an energy source to support the cluster against continued collapse.  Since we stop both models at core collapse, we do not show the post-collapse evolution of either cluster; however, the cluster model with binary physics has its core collapse time delayed by $0.15$ in units of the half-mass relaxation time:

\begin{equation}
T_{\rm rlx} = 0.138 \frac{N}{\log(\gamma N)}\left(\frac{r_h^3}{GM}\right)^{1/2}~,
\end{equation}

\noindent where we have used the more precise definition from \cite{Spitzer1987}, rather than the approximate equation \eqref{eqn:approxtrel}.  Similarly, the number of binaries in the core increases dramatically during the final moments of core collapse, as the central density increases sharply.  In both cases the cluster collapses at about $16.8 T_{\rm rlx}$, consistent with both theoretical predictions and previous numerical integrations \citep{1979ApJ...234.1036C,Cohn1980,1996MNRAS.282...19S,1996NewA....1..255Q,Freitag2001}.

  \begin{figure*}
\centering
\includegraphics[]{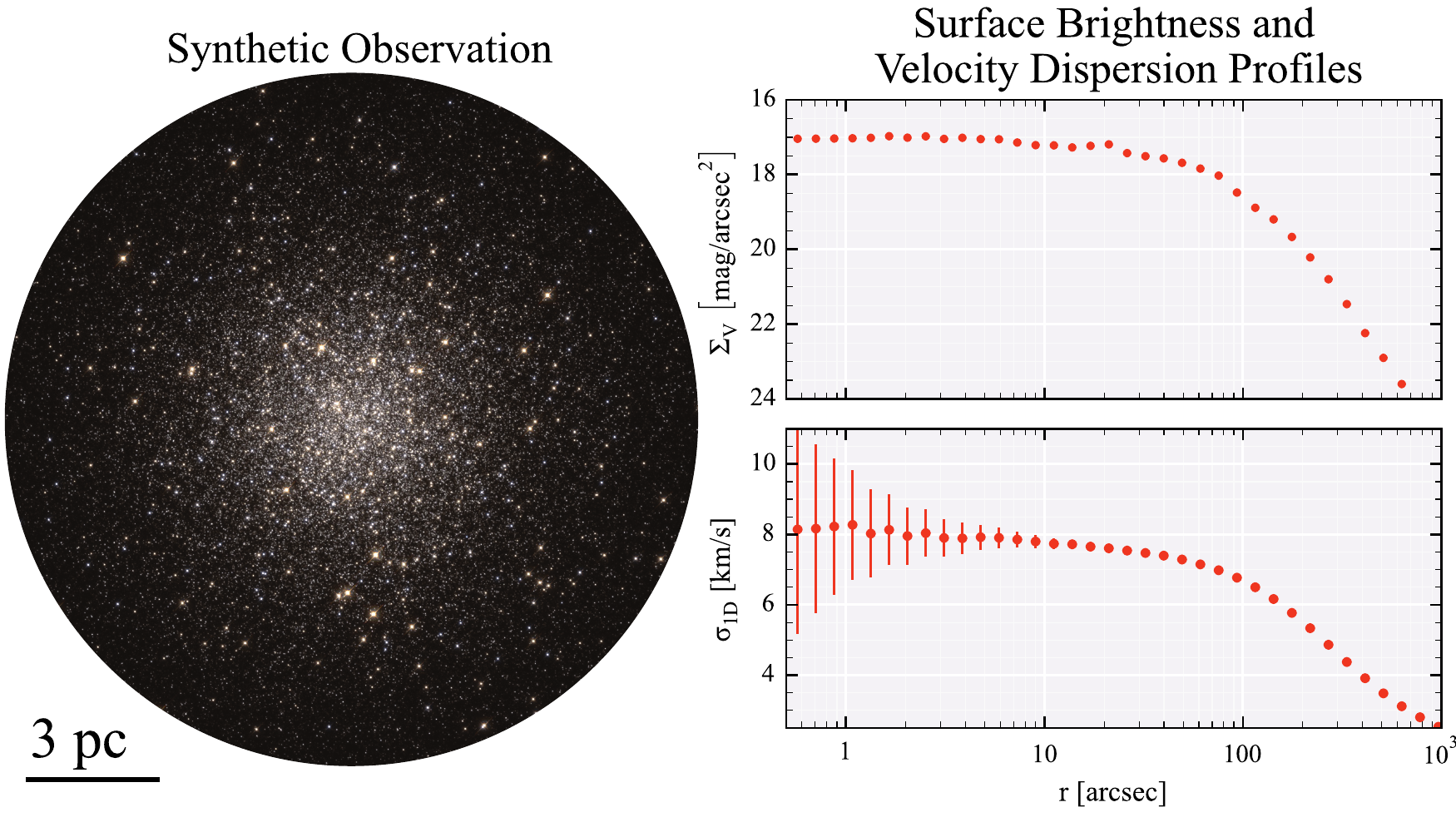}
\caption{An example snapshot of a realistic cluster model after 12 Gyr of evolution.  On the \textbf{left}, we show a synthetic HST observation of the cluster, computing using the stellar positions, effective temperatures, and radii \cite[via the \texttt{Fresco} package,][]{steven_rieder_2019_3553805}.  On the \textbf{right}, we show the V-band surface brightness profile (at a distance of 5 kpc) and 1-D velocity dispersion for the cluster, calculated using the \texttt{cmctoolkit} package \citep{2021zndo...4579950R}.}
\label{fig:pretty}
\end{figure*}

  \begin{figure*}
\centering
\includegraphics[]{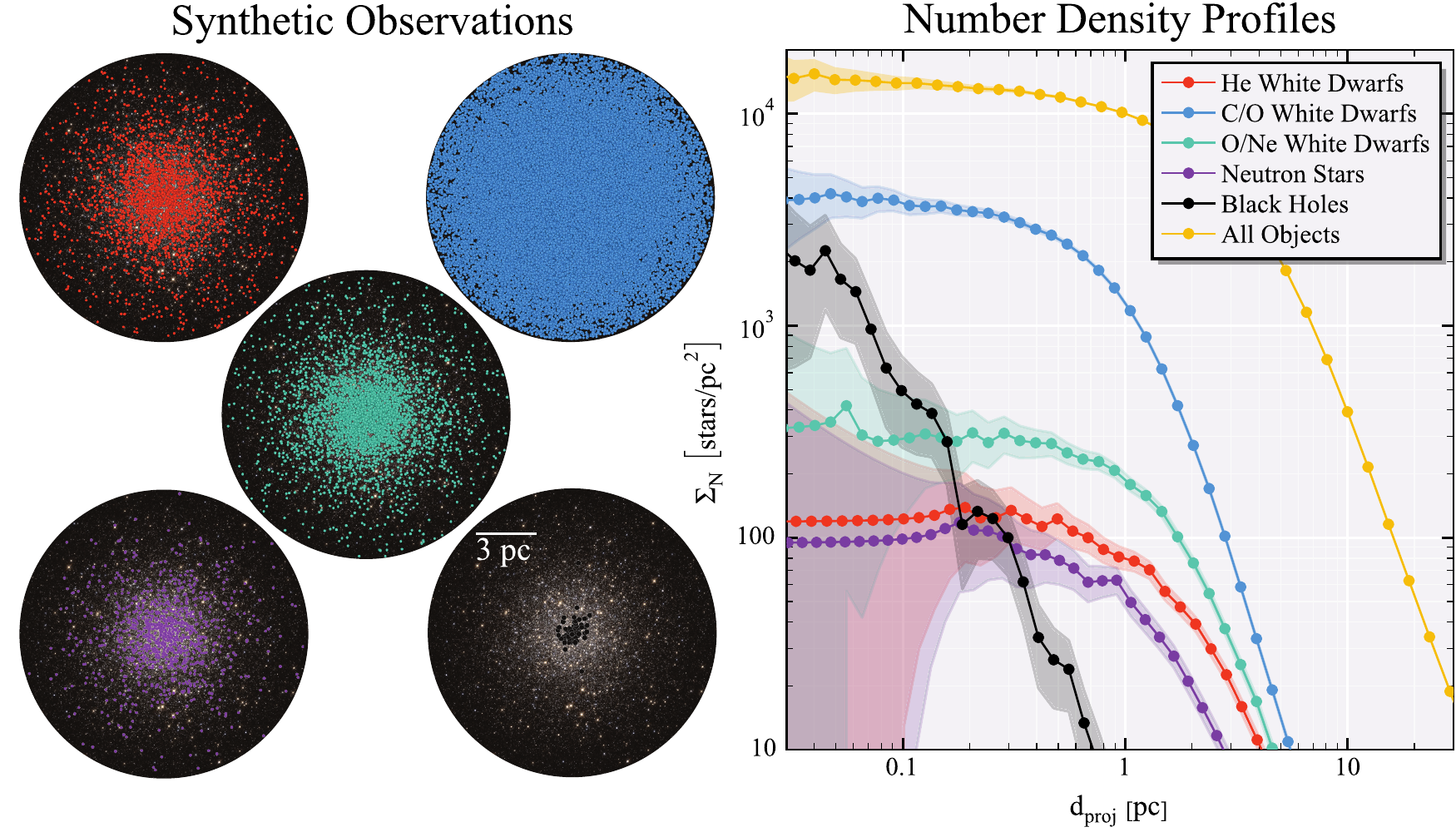}
\caption{The same snapshot from the left of Figure \ref{fig:pretty}, but showing the various populations of low-luminosity particles hidden in the cluster's surface brightness profile.  On the \textbf{left}, we show the distribution of Helium, Carbon/Oxygen, and Oxygen/Neon white dwarfs (in red, blue, and teal) as well as the neutron stars (purple) and BHs (black).  On the \textbf{right}, we show smoothed 2-D number densities and their uncertainties for these populations, calculated from the same snapshot using the \texttt{cmctoolkit}.}
\label{fig:pretty2}
\end{figure*}

It is also well-known that the collapse of a cluster proceeds homologously, with the central density and time to core collapse evolving at a self-similar rate throughout the cluster's central regions  \citep{1961AnAp...24..369H,LyndenBell1975}.   Both homologous models and 1-D Fokker-Planck models suggest that the central density profile of a cluster near collapse should approach a power law of the form $\rho \propto r^{-2.23}$ \citep{LyndenBell1980,Cohn1980,Heggie1988}.  In Figure \ref{fig:rhor}, we show the density of stars (averaged over 20 nearest neighbors) in our cluster model without binary formation as it approaches core collapse.  The density of the central regions increases dramatically as the core approaches final collapse, with the central density increasing by nearly 4 orders of magnitude within the last $10^{-5}~T_{\rm rlx}$ of the evolution.  At the moment of core collapse, \texttt{CMC} reproduces the $r^{-2.23}$ distribution over more than 15  orders of magnitude.  To the best of our knowledge, this is the largest fully-collisional star-by-star model of a Plummer sphere presented to date.

Finally, a critical test of any $N$-body scheme is how well it conserves energy, since this serves as an independent benchmark of the self-consistency of the physics.  By itself, two-body relaxation as implemented in the H\'enon algorithm automatically conserves energy; in practice, with the energy-conserving corrections described in Section~\ref{sec:energyconv}, a \texttt{CMC} run typically conserves energy to within one part in $\sim 10^5$ over many relaxation times.  We show the differential change in energy for both models as a function of time in Figure~\ref{fig:energy}.

\subsection{Realistic Globular Clusters}
\label{s:king}

To demonstrate the capability of \texttt{CMC} to generate realistic cluster models, we generate 10 models of star clusters with random initial conditions drawn from the same distribution function.  We begin by sampling positions and velocities for $10^6$ particles using a \cite{King1966} profile (with $W_0=6$) and a virial radius of $1\,$pc.   The properties for each star and binary are sampled using the same initial condition generators as \texttt{COSMIC}: each star is first assigned a mass from a \cite{Kroupa2001} initial mass function (IMF).  We randomly assign 10\% of those stars to be binaries, keeping the initial IMF draw as the primary mass, and assigning a secondary mass from a uniform distribution between 0.1 and 1 times the primary mass.  For isolated binaries, the ``independent'' sampler in \texttt{COSMIC} assigns binary semi-major axes from any number of user-specified distributions, e.g., flat in the log-period \citep{1983ARA&A..21..343A}, or the distributions from \cite{Sana2012} and \cite{Renzo2019}.  However, cluster environments come with an additional complication: many wide binaries that are observed in galactic fields (and are reproduced by the distributions in \texttt{COSMIC}) are significantly larger than the typical inter-particle separation of the cluster, and would be very rapidly destroyed by dynamical encounters (or simply not formed at all).  To avoid this, we truncate the orbital periods at the local hard-soft boundary for {a binary with the given mass at }that position in the cluster (this approximately corresponds to limiting the minimum orbital velocity to the local three-dimensional velocity dispersion).  For the binaries considered here, we assume a flat-in-log distribution of semi-major axes, and a thermal distribution of eccentricities \citep{Ambartsumian1937,Heggie1975}.  We set stellar metallicities equal to $Z = 0.0002$ (approximately 1\% of the solar value), and do not place the clusters in an external galactic potential.  

In Figure \ref{fig:pretty}, we show a projected snapshot of one of the cluster models rendered using the \texttt{fresco} package \citep{steven_rieder_2019_3553805}, which incorporates the stellar positions, effective temperatures, and radii to construct a mock HST observation.  We also show the surface brightness and velocity dispersion profiles for that snapshot, calculated using the \texttt{cmctoolkit} \citep{2021zndo...4579950R,2021arXiv210305033R}.  The toolkit can directly convert the HDF5 output snapshots of \texttt{CMC} into observational quantities, including surface brightness and velocity dispersion profiles, number density profiles, counts of blue stragglers and other stellar types, and so on.  Combined with the public release of \texttt{CMC}, these two software packages contain all the necessary tools to create models of GCs and SSCs that can be directly compared to observations in the local universe.

One of the main advantages of $N$-body modeling is the ability to study the internal dynamics of lower-luminosity components of the cluster.  In Figure \ref{fig:pretty2}, we show the populations of white dwarfs (WDs), neutron stars (NSs) and BHs that are present throughout the cluster model shown in Figure \ref{fig:pretty} (each overlaid over the synthetic observation), as well as the projected and smoothed 2-D number densities as calculated with the \texttt{cmctoolkit}.  As expected, the heavier components---the Carbon-Oxygen and Oxygen-Neon WDs, NSs, and BHs---have segregated into the central regions of the cluster.  The BHs, in particular, are highly concentrated, with 50\% of the BHs lying within 0.2 pc of the cluster center, where the number density of BHs is more than $7000~\rm{pc}^{-3}$.

In a realistic GC, massive stars evolve in $\sim10-100\,\myr$ after cluster formation, leaving behind a large population of BH remnants.  These BHs rapidly sink to the center, where they begin to participate in the three- and four-body interactions that form and dynamically harden binaries. This process continues without limit until all the BH binaries are either ejected from the cluster (having each also ejected several single BHs during three-body encounters) or merge due to GW emission. The rate of BH depletion is determined by the energy required to support the cluster against continued collapse \cite[e.g.,][]{Breen2013}, and in clusters with sufficiently small initial virial radii, this process can completely remove the BH subsystem from the cluster, producing the core-collapsed clusters observed in the MW \citep{Kremer2019a}.  Overall, the dynamics of BH subsystems are expected to play a key role in shaping the overall evolution of DSCs \citep[e.g.,][]{Merritt2004,Mackey2007,Mackey2008,BreenHeggie2013,Wang2016,Chatterjee2017a,ArcaSedda2018,Kremer2018b,Kremer2020a,Kremer2020,Antonini2020,Weatherford2021}.  The particular cluster simulation shown in Figure~\ref{fig:pretty} still retains 125 BHs by the present day, producing the large core radius observed in the surface brightness profile.

\subsection{Binary Black Hole Mergers and GW190521}
\label{subsec:bbh}

  \begin{figure}
\centering
\includegraphics[]{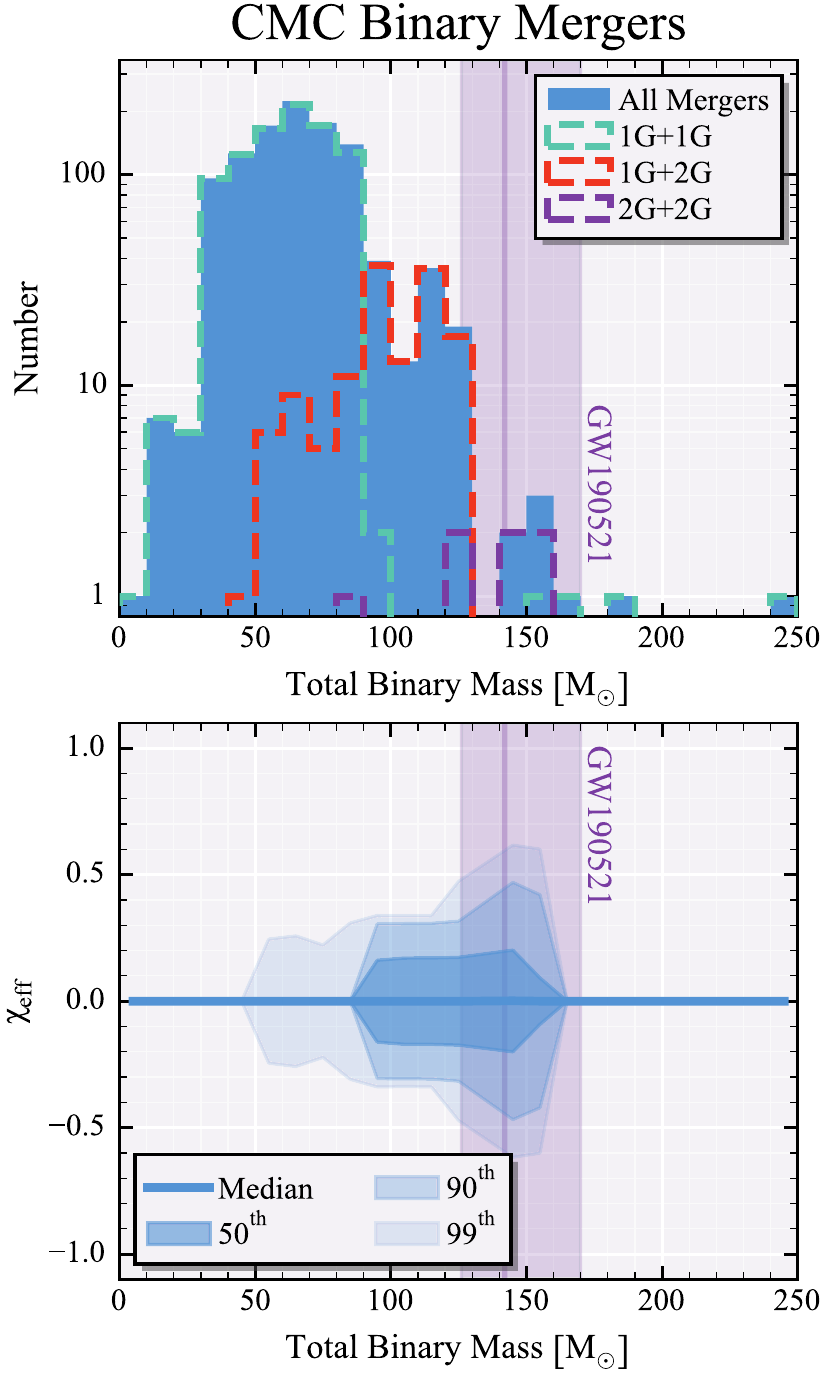}
\caption{The masses and effective spins for merging BBHs from 10 realizations of a  GC model with $10^6$ initial particles (10\% of which are binaries).  On the \textbf{top}, we show the distribution of total masses of all BBHs that merge within 13.8~Gyr of cluster formation.  We also indicate the specific generation of BHs involved in the mergers with the dashed histograms.  The region consistent with GW190521 is populated by 2G+2G BBHs (where both BHs were formed in previous mergers) and 1G+1G BBHs whose component progenitors participated in previous stellar mergers.  On the \textbf{bottom}, we show the distribution of effective spins for the full BBH population as a function of mass.  At higher mass, the contribution from 2G BHs increases, and the range of detectable spins increases as well.}
\label{fig:bbh}
\end{figure}

  \begin{figure}
\centering
\includegraphics[]{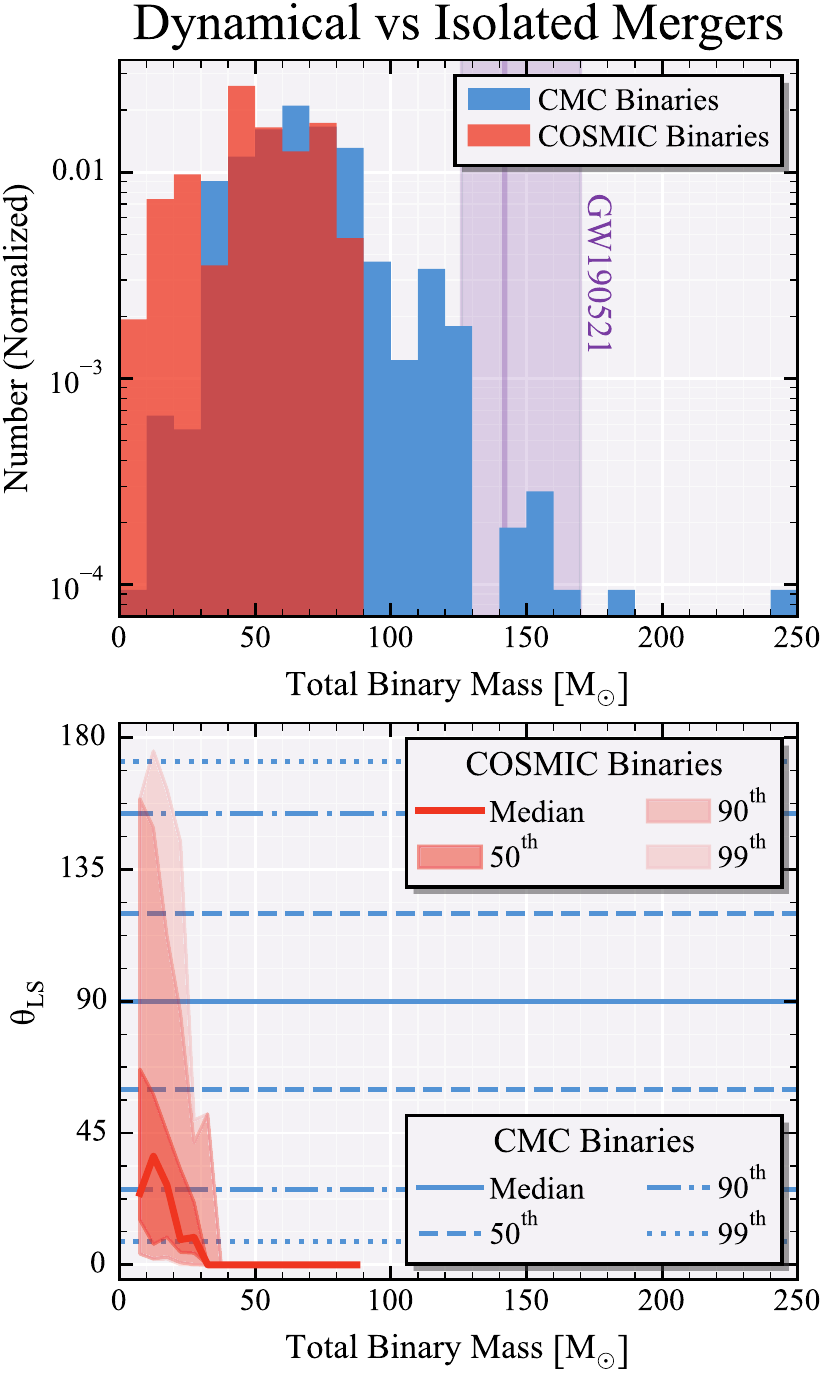}
\caption{Masses and spin alignments of BBHs produced through dynamical encounters (from 10 GC models generated with \texttt{CMC}) and through isolated binary evolution (from $5\times10^4$ stellar binaries evolved with \texttt{COSMIC}).  On the \textbf{top}, we show the total masses of the BBHs, with the maximum mass from isolated binaries ($89M_{\odot}$) being clearly visible.  On the \textbf{bottom}, we show the distribution of spin-orbit misalignments for the BBHs as a function of total mass.  While low-mass BBHs from the isolated binary population can experience significant misalignments from supernova natal kicks, this decreases as a function of mass, while the spin-orbit misalignments for dynamically-formed binaries are isotropic irrespective of mass.}
\label{fig:cmcvscosmic}
\end{figure}
\

In addition to the global dynamical and observable properties of present-day clusters, \texttt{CMC} is ideal for understanding the rate and properties of many high-energy transients and compact binary sources that originate in DSCs.  As previously stated, the BHs in the central regions of a cluster will continually form binaries and undergo hardening encounters, providing a finite energy source to support the cluster against gravothermal collapse.  While this behavior is critical to understanding the overall evolution of the cluster and its present-day appearance, it also produces a large number of BBHs, and is thought to provide a key formation channel for the GW sources detected by LIGO, Virgo, and KAGRA (LVK).  This process has been studied theoretically for many years \citep{Kulkarni1993,Sigurdsson1993,PortegiesZwart2000,OLeary2007,Downing2010,Downing2011,Aarseth2012,Tanikawa2013,Bae2014,Askar2016,FragioneKocsis2018}, including in several papers using \texttt{CMC} \cite[e.g.,][]{Rodriguez2015a,Rodriguez2016a,Rodriguez2016b,Chatterjee2017b,Chatterjee2017a,Rodriguez2018c,Rodriguez2018,kremer2019,rodriguez2019,Kremer2020,Kremer2020b,2020ApJ...896L..10R,2020ApJ...903...67M,Holgado2021,Weatherford2021,Gonzalez2021,rodriguez2021,2021arXiv210501671M}.  

With over 50 GW detections so far, much work has been done trying to determine the origin of the observed population of merging BBHs.  One well-studied technique to discriminate between BBHs formed dynamically in clusters and those formed from isolated binary evolution in the field is to measure the spins of the BHs \cite[e.g.,][]{Rodriguez2016c,Farr2017,Farr2018}, since dynamically-formed BBHs are expected to have isotropically distributed spin-orbit misalignments (versus the relatively small spin-orbit misalignments expected from binary star evolution).  Using the distribution of BBH spins, the latest catalog from the LVK suggests that between 25\% and 93\% of these systems may have been assembled dynamically \citep{2020arXiv201014533T}, with more sophisticated analyses of BBH formation models (including those created using \texttt{CMC}) reaching similar conclusions \citep{Zevin2021,Wong2021}. 

In addition to statistical evidence, individual BBH events have also suggested a dynamical origin.  One particular system, GW190521, was detected with at least one component mass inside the ``upper mass gap,'' where individual BHs are prevented from forming by stellar collapse due to pulsational-pair instabilities, which eject large amounts of material from (and potentially destroy) any star with a helium core mass between about $40\,M_{\odot}$ and $130\,M_{\odot}$ \cite[e.g.,][]{Woosley2017}.   While it is extremely difficult (though not impossible) to describe such systems as the product of isolated stellar evolution \cite[e.g.,][]{Farmer2019,Belczynski2020}, these systems can be easily produced in dense star clusters, though either the repeated mergers of BHs \cite[e.g.,][]{Rodriguez2018,rodriguez2019,FragioneLoeb2020,FragioneSilk2020} or through massive star mergers occurring prior to BH formation \cite[e.g.,][]{DiCarlo2019,Kremer2020b,Weatherford2021,Gonzalez2021}.  

Figure \ref{fig:bbh} shows the total masses and spins of all BBHs that merge within 13.8~Gyr of cluster formation from our 10 GC models.  We subdivide the population according to the ``generation'' of the BHs in the binary, with first generation, or 1G, BHs being those created from stellar collapse, and second generation, or 2G, referring to those created in a previous merger.  The more massive BHs, particularly those within the mass range consistent with GW190521, are predominantly the result of previous mergers, with 4 of the 6 systems being comprised of 2G+2G BBHs (that is, binaries whose components are \emph{both} the product of previous mergers).  The remaining two BBHs are created from previous \emph{stellar} mergers of massive stars prior to their collapse into BHs.  These stellar merger products can contain unique envelope/core mass ratios, allowing them to bypass the typical pulsational-pair instability limits for single massive stars.  While repeated BH mergers are the dominant mechanism for producing GW190521-like systems in these models, cluster with higher central densities, fractal initial conditions, or different stellar binary fractions can prefer formation of repeated mergers through stellar collisions \cite[e.g.,][]{DiCarlo2019,Kremer2020b,Gonzalez2021}.  {We note that while our models also ignore the possibility of accretion from BH/star collisions, the majority of these occur after the most massive stars (including stellar-merger products) have expired, since the BHs typically require $\sim 100$ Myr to dynamically segregate into the cluster center.  As such, $M_{\rm BH} > M_{\rm star}$ for the majority of BH/star collisions.  }

We also show in Figure \ref{fig:bbh} the distribution of the effective spins of the cluster BBH population as a function of the mass.   This mass-weighted projection of the BH spins onto the orbital angular momentum, $\chi_{\rm eff}\equiv \hat{L}\cdot\left[(m_1\vec{\chi_1} + m_2\vec{\chi_2})\big/(m_1+m_2)\right]$, is the spin parameter most easily constrained by GW detectors \citep[e.g.,][]{2011PhRvL.106x1101A,Abbott2016e}. Since we have assumed that BHs from collapsing stars are born with zero spin, consistent with recent theoretical studies of angular momentum transport in massive stars \citep[e.g.,][]{2019MNRAS.485.3661F,2019ApJ...881L...1F}, BBHs whose components are formed from stellar collapse also have zero effective spins across all masses.  On the other hand, when BHs are created from previous mergers, their spins inherit the angular momenta of their parent binaries, producing a population of spinning BBHs.  Because the BH spins are assumed to be isotropically distributed on the sphere, the median effective spin for all cluster BBHs is still centered at $\chi_{\rm eff} = 0$.  This is consistent with the observation of GW190521, which appears to be consistent with having zero effective spin.  However, as consistent with our previous results \cite[e.g.,][Figure 3]{Rodriguez2018}, any \emph{population} of 2G+2G BBHs (like the ones that predominately form GW190521-like binaries here) should have a range of effective spins centered at $\chi_{\rm eff}=0$.

As a demonstration of the utility of the \texttt{CMC}/\texttt{COSMIC} integration, we directly compare the BBH population created through dynamical processes to that created from the evolution of isolated stellar binaries.  We evolve a population of $5\times10^4$ binaries with masses $\geq 18M_{\odot}$ using the same stellar physics, stellar metallicitiy, and binary initial conditions (the truncation in binary orbital period) with \texttt{COSMIC}.    In Figure~\ref{fig:bbh}, we show one of the most distinct features of the BBHs produced by GCs: the production of repeated mergers of BHs in star clusters.  When two BHs merge in a GC, their merger product receives a recoil kick due to the asymmetric emission of GWs, which depends on both the mass ratio of the system and the spins of the BHs \cite[e.g.,][]{Merritt2004,Lousto2008,Campanelli2007}.  Because three-body encounters preferentially form BBHs with near-equal mass components \cite[e.g.,][]{Sigurdsson1993a}, it is the BH spins that primarily determine merger retention in GCs.  While large spins ($\chi \sim 1$) leave few second generation BHs in the star cluster, mergers of BHs born with smaller spins ($\chi \lesssim 0.2$) can be retained in many GCs, where they will continue to participate in dynamical processes, form binaries, and merge again \citep{rodriguez2019}.  This process can produce a significant population of BBHs with large ($\chi \sim 0.7$) spins and masses in the pulsational-pair instability mass gap.

While an observation of a BBH with $\chi_{\rm eff} < 0$ is strongly suggestive of dynamical formation, isolated binary evolution can also produce BBHs with spin-orbit tilts greater than $90^{\circ}$ given the right orientation and magnitude of the supernova kicks \citep{Kalogera2000,Rodriguez2016c}.  Therefore, any analysis of BBH spin-orbit misalignments must include both dynamical and binary evolutionary channels using identical stellar physics.  The integration of \texttt{COSMIC} into \texttt{CMC} makes this trivial; in Figure \ref{fig:cmcvscosmic}, we show the masses and spin-orbit misalignments ($\theta_{\rm LS}$) from the 10 GC models and $5\times10^4$ isolated massive binaries evolved using \texttt{COSMIC}.  In the top panel, the limit imposed by pulsational-pair instability physics on the BH mass function is readily apparent, with no BHs forming with masses greater than $44.5M_{\odot}$, and no BBHs forming with total masses greater than $89M_{\odot}$ \cite[a number that is largely insensitive to choices of accretion and other binary physics,][]{2020ApJ...897..100V}. Figure \ref{fig:cmcvscosmic} also clearly demonstrates the difference in the spin-orbit misalignments between isolated binaries and cluster-formed BBHs.  Given the supernova prescriptions in \texttt{COSMIC} \cite[here we have used the ``rapid'' prescription from][]{Fryer2012}, BHs with final core masses $\gtrsim 20M_{\odot}$ form via the ``direct collapse'' or failed supernova scenario, where the BHs form without any natal kick or change to the supernova orbital parameters.  For BBHs with total masses $\lesssim 45M_{\odot}$, the supernova kicks can still produce significant spin-orbit misalignments, with the $99^{\rm th}$ percentile of allowed misalignments approaching complete anti-alignment ($\theta_{\rm LS} = 180^{\circ}$).  However, as the mass increases, the amount of spin-orbit misalignment decreases, until eventually all BBHs are aligned with their orbital angular momenta.  Contrast this with BBHs formed in clusters, which are isotropically distributed on the sphere (i.e.~$P(\theta_{\rm LS})d\theta_{\rm LS}\propto \sin(\theta_{\rm LS})d\theta_{\rm LS}$).  These results can be easily reproduced and expanded upon using the apples-to-apples comparisons enabled by the \texttt{CMC} and \texttt{COSMIC} integration.

\
\
\section{Discussion and Conclusion} \label{S:summary}

This paper describes the first public release of the Cluster Monte Carlo code, \texttt{CMC}, a parallel code for collisional $N$-body stellar dynamics based upon the original method of \cite{Henon1971a,Henon1971b}.  After nearly two decades of development, \texttt{CMC} contains all of the necessary physics for modeling the evolution of DSCs, such as GCs, SSCs, and certain NSCs, and it allows the user to easily create detailed models of spherical star clusters with up to $\sim 10^7$ particles (when including stellar evolution).  In addition to two-body relaxation, \texttt{CMC} can model the formation and dynamics of binaries, the effects of tidal stripping in a galactic field, detailed binary star evolution, and more.  

The public release of \texttt{CMC} has been coupled to the \texttt{COSMIC} software package for binary population synthesis \citep{Breivik2020}.  The newest release of \texttt{COSMIC}, v3.4, contains new initial condition generators to create cluster profiles from various spherical stellar distributions \citep{Plummer1911,King1966,Elson1987}, which can then be coupled to all of the stellar and binary initial condition generators in the population synthesis code.  Both the initial conditions and the output snapshots of \texttt{CMC} are saved in easily-readable HDF5 format files, and can be analyzed with standard data science packages (pandas) or with specialty software designed to compare \texttt{CMC} models directly to observations of real star clusters \cite[the \texttt{cmctoolkit},][]{2021zndo...4579950R,2021arXiv210305033R}.

The synergy between \texttt{CMC} and \texttt{COSMIC} enables stellar dynamics and population synthesis studies to be conducted simultaneously with identical assumptions and prescriptions for stellar initial conditions.  We presented two examples of \texttt{CMC} in action.  First, we studied the collapse of a Plummer sphere with $10^8$ initial particles, the largest fully collisional star-by-star model integrated to core collapse, and showed that \texttt{CMC} was able to replicate the self-similar nature of the collapse across more than 15 orders of magnitude.  Then we presented examples of realistic clusters and showed how the combination of \texttt{CMC} with \texttt{COSMIC} makes comparative population studies---such as our study of GW190521, a BBH that was very likely formed in a dynamical environment---trivial.  It is our hope that the public release of \texttt{CMC} and the integration of \texttt{COSMIC} will enable significant advancements in the study of DSCs and the many transient and high-energy events they produce (such as BBH mergers).

\
\
\
\
\
\
\acknowledgements

We thank Kuldeep Sharma, Xiaoqi Yu, and Mike Grudi\'c for testing this release of \texttt{CMC} and providing feedback, and Elena Gonz\'alez and Miguel Martinez for useful comments and discussions.  This work was supported by NSF Grant AST-2009916 at Carnegie Mellon University, a New Investigator Research Grant to C.R.~from the Charles E.~Kaufman Foundation, and NSF Grant AST-1716762 at Northwestern University.   This work used the Extreme Science and Engineering Discovery Environment (XSEDE), which is supported by National Science Foundation grant number ACI-1548562. Specifically, it used the Bridges-2 system, which is supported by NSF award number ACI-1928147, at the Pittsburgh Supercomputing Center (PSC).  N.C.W.\ acknowledges support from the CIERA Riedel Family Graduate Fellowship. F.K.\ acknowledges support from the Turkish Fulbright Commission. K.K.\ is supported by an NSF Astronomy and Astrophysics Postdoctoral Fellowship under award AST-2001751. P.A-S.\  acknowledges support from the Ram{\'o}n y Cajal Programme of the Ministry
of Economy, Industry and Competitiveness of Spain, as well as the financial
support of Programa Estatal de Generación de Conocimiento (ref.
PGC2018-096663-B-C43) (MCIU/FEDER). 
N.Z.R.\ acknowledges support from the Dominic Orr Graduate Fellowship at Caltech.\\\\

\software{The public release of \texttt{CMC} can be accessed, including source code and documentation, at this URL: \url{https://clustermontecarlo.github.io/}.  \texttt{CMC} \citep[][this work]{Joshi2000,2000ApJ...539..331W,Joshi2001a,2002ApJ...570..171F,Fregeau2003,Chatterjee2010,Morscher2013,Pattabiraman2013, codepaper2021}, \texttt{cmctoolkit} \citep{2021zndo...4579950R,2021arXiv210305033R}, \texttt{fewbody} \citep{Fregeau2007,Antognini2014,Amaro-Seoane2016}, \texttt{COSMIC} \citep{Breivik2020,2020zndo...3905335B}, \texttt{matplotlib} \citep{Hunter2007}, \texttt{SciPy} \citep{scipy}, \texttt{NumPy} \citep{2020NumPy-Array}, \texttt{pandas} \citep{mckinney-proc-scipy-2010,jeff_reback_2021_4681666}.}

\bibliography{cmcrelease}

\end{document}